\documentclass[10pt,
 notitlepage
 preprint, 
superscriptaddress,
nofootinbib,
 amsmath,amssymb,
 aps, physrev,
]{revtex4-2}
\usepackage{xcolor}
\usepackage{amsmath}
\usepackage{amssymb}

\usepackage{float}
\usepackage{graphicx}
\usepackage{dcolumn}
\usepackage{bm}
\usepackage{soul}

\usepackage[colorlinks=true,
            linkcolor=blue,
            citecolor=blue,
            urlcolor=blue]{hyperref}

\begin{document}

\title{\textbf{Testing the Pseudo-Dirac Nature of Atmospheric Neutrinos} 
}%

\author{Olivia M. Bitter}
\affiliation{Northwestern University, Department of Physics \& Astronomy,
2145 Sheridan Road, Evanston, IL 60208, USA}
\author{Andr\'e de Gouv\^ea}
\affiliation{Northwestern University, Department of Physics \& Astronomy,
2145 Sheridan Road, Evanston, IL 60208, USA}
\author{Srubabati Goswami}
\affiliation{Physical Research Laboratory, Ahmedabad, India, 380009
}%
\author{Supriya Pan}
\affiliation{%
Universidad Católica del Norte, Antofagasta, Chile, 1240000}%
\affiliation{Physical Research Laboratory, Ahmedabad, India, 380009
}


\begin{abstract}
Pseudo-Dirac neutrinos are characterized by extremely small mass-squared differences between two nearly degenerate, strongly-mixed, active--sterile states. These differences are associated with new very-long oscillation lengths and provide a probe of the nature of neutrinos. Atmospheric neutrinos can probe these oscillations as they allow access to a broad range of baseline distances, from tens of kilometers to the Earth’s diameter. We investigate the sensitivity of an experiment like the Deep Underground Neutrino Experiment (DUNE) to the pseudo-Dirac nature of neutrinos through measurements of the atmospheric neutrino flux. We concentrate on the modifications to the atmospheric neutrino energy and angular spectra induced by one new oscillation frequency associated with the $\nu_3$ state and determine the parameter space that can be probed with atmospheric neutrinos at a DUNE-like detector.  We also comment on the sensitivity of these measurements to the sub-leading ''solar'' oscillations, which are also characterized by a relatively small mass-squared difference. 

\end{abstract}


\maketitle

\section{Introduction}

The discovery of neutrino oscillations in solar, atmospheric, accelerator, and reactor experiments embodies one of the most important breakthroughs in particle physics over the past few decades. It constitutes the first direct evidence of physics beyond the Standard Model (SM), since neutrinos are massless in the original formulation of the SM.

Neutrino oscillations are quantum mechanical interference processes in which neutrinos change flavors after traversing large distances. 
The flavor-conversion probability depends on the propagation distance $L$, the neutrino energy $E$, the mass-squared differences $\Delta m_{ij}^{2}=m_i^2-m_j^2$ among the neutrinos and mixing angles $\theta_{ij}$ ($i,j=1,2,3, i\neq j$) that characterize charged-current weak interactions \cite{ParticleDataGroup:2026mpi}.
The current data have established the three-flavor paradigm of oscillation on a rather firm footing, and the mixing angles and mass-squared differences are measured with impressive precision. The three main features of the standard picture that remain undetermined are the sign of $\Delta m^2_{31}$, i.e., the neutrino mass ordering, the octant of the mixing angle $\theta_{23}$, and the Dirac CP phase $\delta_{CP}$. 

Neutrino oscillation experiments remain insensitive to the absolute neutrino mass scale. Information about this parameter can come from experiments on beta decay and neutrinoless double beta decay as well as from cosmology. Among these, the combined analysis of DESI DR2 plus Planck CamSpec provides the best bound of $0.0642$ eV on the sum of neutrino masses at the 95\% level, disfavoring the inverted mass ordering \cite{jimenez2026evidenceevidentdecisivecosmological}.
The small mass-squared differences measured in oscillation experiments, together with the cosmological mass bound, imply that the neutrino masses are very small, safely below 0.1~eV. 

Understanding the origin of such small neutrino masses remains one of the central open questions in particle physics. 
Unlike charged fermions, neutrinos are electrically neutral and can be Majorana or Dirac particles. A Dirac neutrino consists of four independent degrees of freedom corresponding to left-handed and right-handed neutrinos and antineutrinos. In this case, total lepton number is a conserved symmetry. Dirac neutrinos require the introduction of right-handed neutrino fields that are singlets under the SM gauge group. 
In contrast, Majorana neutrinos are their own antiparticles. A Majorana neutrino consists of two independent degrees of freedom: a left-handed and a right-handed one. Majorana neutrino masses violate the lepton number by two units. Such terms can arise through the dimension-five Weinberg operator \cite{Weinberg:1979sa}  whose ultraviolet completion might come via one of the various seesaw mechanisms~\cite{Minkowski:1977sc,Yanagida:1979as,GellMann:1979vob,Mohapatra:1979ia,Konetschny:1977bn,Magg:1980ut,Cheng:1980qt,Schechter:1980gr,Lazarides:1980nt,Mohapatra:1980yp,Foot:1988aq} (for an overview, see, for example, \cite{deGouvea:2016qpx}).     
These can be naturally implemented in grand unified theories and provide an elegant formulation for generating small neutrino masses in which the violation of lepton number is related to the small neutrino masses. 
Neutrino oscillations cannot distinguish between the Dirac and Majorana nature of the neutrinos. Evidence for this can come from lepton-number-violating processes such as neutrinoless double beta decay.  Determining the Dirac or Majorana nature of neutrinos is among the primary objectives of particle physics.

An intriguing intermediate possibility is that neutrinos are pseudo-Dirac fermions~\cite{Wolfenstein:1981kw,Petcov:1982ya,Valle:1983dk,Doi:1983wu}. In this scenario, the dominant neutrino mass originates from a Dirac mass term, while small Majorana mass terms split each Dirac neutrino into two nearly degenerate Majorana states. Pseudo-Dirac neutrinos are Majorana fermions but behave, as far as most experimental probes are concerned, as Dirac neutrinos.

As far as oscillations are concerned, if the neutrinos are pseudo-Dirac, every active neutrino is accompanied by a sterile partner and a new, extremely small mass-squared difference that satisfies
\begin{equation}
\Delta m_{\rm PD}^{2}\ll
\Delta m_{21}^{2},
 |\Delta m_{31}^{2}|,
\end{equation}
such that conventional three-flavor oscillation phenomenology remains essentially unchanged at short-enough baselines. However, oscillations into sterile partners are expected to become observable for sufficiently large values of $L/E$. Hence, the presence of pseudo-Dirac neutrinos modifies the standard oscillation framework by introducing additional oscillation frequencies associated with the tiny active--sterile mass splittings,  which changes the flavor transition probabilities \cite{Joshipura:2000ts, Kobayashi:2000md}.   
Assuming the mixing between the active--sterile pair to be maximal, robust constraints have been obtained on the mass splitting associated with the first two mass eigenstates,
$\Delta m^2_{1 1^\prime}, \Delta m^2_{2 2^\prime} \lesssim 10^{-11}$ eV$^2$ \cite{de_Gouv_a_2009, Donini:2011jh, Ansarifard:2022kvy}. Existing constraints on the mass splitting associated with the third mass eigenstate from atmospheric, accelerator, and reactor data are considerably weaker:  $\Delta m^2_{3 3^\prime} \lesssim 10^{-5}$ eV$^2$ \cite{Anamiati_2018}.  
Studies have also been done to probe the pseudo-Dirac nature of neutrinos in future reactor and accelerator facilities  \cite{Anamiati:2019maf, Boudjema:2026caj, abada2025investigatingduneoscillationssensitivity, sen2023constrainingpseudodiracneutrinosgalactic} and a study to probe the pseudo-Dirac nature in JUNO using solar neutrinos can be found in \cite{Franklin:2023diy}. 
Since these effects accumulate only over extremely large baselines, high-energy astrophysical neutrinos provide unique opportunities to probe such scenarios~(see, for example, \cite{PhysRevLett.92.011101, Pakvasa:2012db, Joshipura:2013yba, Rink:2022nvw,Carloni:2022cqz,Dev:2024yrg,Dixit:2024ldv,Carloni:2025dhv,Leal:2025eou,Cabrera:2025fmj,yc27-61w5,MacDonald:2025jbm}).
 
Among all available neutrino sources, atmospheric neutrinos span energies from sub-GeV to multi-GeV while traversing baselines ranging from a few tens of kilometers to the Earth's diameter. Consequently, atmospheric neutrinos probe an enormous range of $L/E$, making them potentially sensitive to tiny mass-squared splittings. Furthermore, the propagation of atmospheric neutrinos through the Earth experiences matter effects that can significantly modify active--sterile oscillations.

In this work, we investigate pseudo-Dirac neutrinos at a DUNE-like detector.  
Motivated by existing experimental constraints, we concentrate on the pseudo-Dirac splitting associated with the third mass eigenstate, $\Delta m_{33'}^{2}$, which presently remains the least constrained among the pseudo-Dirac splittings and is most relevant for the oscillations of atmospheric neutrinos. 
Hence, we investigate pseudo-Dirac neutrinos within the framework of a $3+1$ oscillation formalism. We study the modifications to neutrino oscillation probabilities, investigate their impact on atmospheric neutrinos, and assess the sensitivity of a DUNE-like detector to probe these effects via observations of atmospheric neutrinos.

The paper is organized as follows. Section~\ref{sec:psedudodirac} introduces the pseudo-Dirac neutrino framework together with the relevant theoretical motivation and the $3+1$ oscillation formalism of interest.  Section~\ref{sec:atmospheric_nus} describes the numerical methodology and the atmospheric neutrino analysis. The results and phenomenological implications are discussed in Section~\ref{sec:PD_sensitivity}.  Section~\ref{sec:solar} contains a discussion of how sensitive is the atmospheric neutrino data to "solar" oscillations. Finally, Section~\ref{sec:conclusions} summarizes our results. 

\section{Pseudo-Dirac Neutrinos}
\label{sec:psedudodirac}

We start by postulating that the neutrino masses are a consequence of the existence of gauge-singlet ``right-handed-neutrino'' fields.
In this case, the most general renormalizable Lagrangian consistent with SU(3)$_c$ × SU(2)$_L$ × U(1)$_Y$ gauge invariance is
\begin{equation}
{\cal L}_{\nu\rm SM}={\cal L}_{\rm old}-\lambda_{\alpha i}L^{\alpha}HN^i-\sum_{i=1}^n\frac{M_{i}}{2}N^iN^i + \text{H.c.},
\label{Lagrangian}
\end{equation}
where ${\cal L}_{\rm old}$ is the minimal SM Lagrangian, $H$ is the Higgs weak doublet, $L^{\alpha}$,  $\alpha=e,\mu,\tau$, are lepton weak doublets, $\lambda_{\alpha i}$ are neutrino Yukawa couplings, and $M_i$ are Majorana masses for the right-handed neutrino fields $N_i$ fields that have been added. The number of $N^i$ fields is $n$ and, in order to fit the neutrino oscillation data, $n\ge 2$. Henceforth, when relevant, we assume $n=3$.

After electroweak symmetry breaking, the $6\times 6$ neutrino mass matrix in the $\begin{pmatrix}\nu_L \\N^c\end{pmatrix}$ basis is
\begin{equation}
\mathcal{M}
=
\begin{pmatrix}
0 & m_D \\
m_D^T & M
\end{pmatrix},
\end{equation}
where $m_D=\lambda v$ is the usual Dirac mass related to the Higgs vacuum expectation value $v$. 
 $M = 0$ yields the Dirac limit (i.e., pairs of mass-degenerate Majorana fermions form Dirac fermions). On the other hand, when $M \gg m_D$, we can integrate out the right-handed neutrinos and end up with light active Majorana neutrinos (i.e., the seesaw mechanism). 

The limit $M \ll m_D$ yields the pseudo-Dirac case. The neutrino masses are pairwise quasi-degenerate with magnitudes driven by the eigenvalues of $m_D$. The small mass splittings are proportional to $M$. We refer to these as $\Delta m^2_{ii'}$, $i=1,2,3$, where $\nu_i'$ and $\nu_i$ are the six different pairwise-quasi-degenerate states. 

Current constraints from solar neutrinos have greatly constrained two out of the three new oscillation frequencies $\Delta m^2_{ii'}$ associated with the $\nu_1$ and $\nu_2$ states \cite{de_Gouv_a_2009, Donini:2011jh, Ansarifard:2022kvy, Anamiati_2018, Franklin:2023diy}. Since solar neutrino experiments primarily involve electron neutrinos that have negligible mixing with $\nu_3$, solar data provides comparatively weaker constraints on $\Delta m^2_{33'}$. Hence, bounds from Earth-bound experiments can compete with those from solar neutrinos in spite of the very long Earth-Sun distance. 

$\Delta m^2_{33'}$ is proportional to $m_3^2$ (see, for example, \cite{de_Gouv_a_2009}), where $m_3$ is the average mass of $\nu_3$ and $\nu_3'$. In turn, for the normal neutrino mass ordering and ignoring the possibility that all neutrino masses are quasi-degenerate, $m_3^2$ is of order $\Delta m^2_{31}$, the so-called atmospheric mass-squared difference. Hence, we define the relevant new oscillation frequency via 
\begin{equation}
\Delta m^2_{33'} = 2 \epsilon \Delta m^2_{31}, 
\label{PD_parm}
\end{equation}
where $\epsilon$ is a dimensionless quantity. Throughout we use $\Delta m^2_{33'}=5\times10^{-3}\times \epsilon$ eV$^2$. We will further assume that the active--sterile mixing angle is maximal, as will become clearer in the next subsection.\footnote{The case discussed in \cite{Wolfenstein:1981kw} was more general and allowed for non-maximal mixing. 
Phenomenology of such cases has been discussed in detail in \cite{Joshipura:2000ts}.} We refer to \cite{de_Gouv_a_2009} for more details. 

\subsection{The pseudo-Dirac \texorpdfstring{$3+1$}{3+1} oscillation formalism}

In the limit $\Delta m^2_{22'}=\Delta m^2_{11'}=0$, the relevant $4\times 4$ Hamiltonian that describes neutrino flavor evolution in the presence of matter is, 
\begin{equation}
     \mathcal{H}_{3+1} = \frac{1}{2 \rm E} \left[U M^2
    U^{\dagger}  + \hspace{1mm} \mathcal {H}_{\rm mat}\right],
\end{equation}
where 
\begin{equation}
    U =  \begin{pmatrix} U_{e 1} &  U_{e 2} & \frac{ U_{e 3}}{\sqrt{2}} & \frac{ -U_{e 3}}{\sqrt{2}}\\ U_{\mu 1} &  U_{\mu 2} & \frac{U_{\mu 3}}{\sqrt{2}} & \frac{-U_{\mu 3}}{\sqrt{2}}\\ U_{\tau 1} &  U_{\tau 2} & \frac{ U_{\tau 3}}{\sqrt{2}} & \frac{ -U_{\tau 3}}{\sqrt{2}}\\ 0 & 0 & \frac{1}{\sqrt{2}} & \frac{1}{\sqrt{2}} \end{pmatrix}.
\end{equation}
$U_{\alpha i}$ are the elements of the standard PMNS mixing matrix, $U_{\rm PMNS}= R_{23}(\theta_{23})R_{13}(\theta_{13})R_{12}(\theta_{12})$,
\begin{equation}
    M^2 =  \begin{pmatrix} 0 &  0 & 0 & 0\\ 0 & \Delta m_{2 1}^2 & 0 & 0\\ 0 &  0 & \Delta m_{3 1}^2 (1 + \epsilon)  & 0\\ 0 & 0 & 0 & \Delta m_{3 1}^2 (1 -\epsilon)  \end{pmatrix},
\end{equation}
and 
\begin{equation}     
\mathcal{H}_{\rm mat} =  \begin{pmatrix} 
A &  0 & 0 & 0\\ 
0 &  0 & 0 & 0\\
0 &  0 & 0 & 0\\
0 &  0 & 0 & \frac{A}{2}\\ 
\end{pmatrix} ,
\end{equation}
where $A=2E V_{\rm CC}$ is the effective contribution in the Hamiltonian due to the matter potential $V_{\rm CC}=\sqrt{2}G_F N_e$, with $N_e$ being the ambient electron density. 
We briefly discuss approximate solutions to the neutrino oscillation probabilities in the presence of constant matter in Appendix~\ref{appA}. 

\section{Atmospheric Neutrinos in a Liquid Argon Detector}
\label{sec:atmospheric_nus}

In modeling our simulations of atmospheric neutrinos, we obtained the atmospheric neutrino flux from the publicly available Honda files using the solar minimum variant for the Homestake site \cite{PhysRevD.92.023004}. Information regarding the total cross sections was obtained through GLoBES \cite{Huber:2007em}, adjusting for $^{40}$Ar as the target of a DUNE-like setup. To account for the matter effects that neutrinos experience along the different trajectories they take through the layers of the Earth from the atmosphere to the detector, we utilized the Preliminary Reference Earth Model (PREM) using an approximation with 5 layers: crust, upper mantle, lower mantle, outer core, and inner core \cite{DZIEWONSKI1981297}.

We consider a setup similar to that proposed for DUNE \cite{DUNE:2020lwj}. DUNE is a long-baseline neutrino experiment under construction in South Dakota, with its far detector located approximately 1.5 km underground. The experiment will ultimately employ a 40 kt liquid argon time projection chamber (LArTPC) to detect neutrinos produced at Fermilab and traversing a baseline of 1300 km. In addition to its primary long-baseline neutrino program, the detector will also be sensitive to atmospheric, astrophysical, and solar neutrinos. The large detector mass, together with its excellent energy and angular resolution, makes this DUNE-like configuration well suited for the study presented here.

The analysis of atmospheric neutrino events is performed using an in-house code. The events per energy and $\cos\theta$ bin are calculated using
\begin{equation}
    \frac{dN}{dE_m d\cos\theta_m}= \iint d(\cos\theta_t)dE_t \phi(E_t, \cos\theta_t) R_E(E_t, E_m) R_\theta(\theta_t, \theta_m) P_{\alpha\beta}(E_t) \phi(E_t, \theta_t) \Sigma(E_t).
\end{equation}
Here $\phi(E_t, \cos\theta_t)$ is the neutrino flux as a function of the energy and direction, and $\Sigma(E_t)$ is the cross section. We consider the energy range of $E_t \in [0.2,10.5]$ GeV and $\cos\theta_t\in[-1,1]$ with 80 bins each. The energy resolutions used for detected electrons and muons are,
\begin{equation}
    \sigma_E(\mu)= 0.1\sqrt{E_t}\text{,   }\sigma_E(e)= 0.2\sqrt{E_t}.
\end{equation}
For the angular resolution, we consider
\begin{eqnarray}
    \sigma_\theta=15 \pi / 180.0 \text{, for } E_t>1~\text{GeV},\\
    \sigma_\theta=\vartheta \pi / 180.0 \text{, for } E_t<1~\text{GeV},
\end{eqnarray}
where we consider $\vartheta$ (in degrees) values equal to $15, 30, 40, 50$. The values for $\sigma_E(\mu),\sigma_E(e), \sigma_\theta(E_\nu >1~\rm{GeV})$ are consistent with those published by the DUNE collaboration in \cite{2026}. For $\sigma_\theta(E_\nu <1~\rm{GeV})$, the resolution in \cite{2026} is most consistent with $\vartheta=50$.  The Gaussian resolution functions are defined as
\begin{align}
R_\theta(\theta_t, \theta_m) & = N_\theta~ \rm{exp}\Big[\frac{\cos\theta_t-\cos\theta_m}{\sqrt{2}\sigma_\theta \sin\theta_t}\Big]^2,\\
 R_E(E_t, E_m) & = N_E ~ \rm{exp}\Big[\frac{E_t-E_m}{\sqrt{2}\sigma_E}\Big]^2,
\end{align}
where the normalization constants are
\begin{align}
N_\theta & = \int_{-1}^{+1} \rm{exp}\Big[\frac{\theta_t-\theta_m}{\sqrt{2}\sigma_\theta}\Big]^2 d\cos\theta_m,\\
N_E & = \int_{E_{min}}^{E_{max}} \rm{exp}\Big[\frac{E_t-E_m}{\sqrt{2}\sigma_E}\Big]^2 d E_m.
\end{align}
We have used 20 bins each for measured energy and $\cos\theta$, with the corresponding ranges being the same as the true case.

We used the events per energy and angular bins for true (ex) and test (th) cases to calculate $\chi^2$ defined as,
\begin{eqnarray}
    \chi^2 & = & \chi_{\rm stat}^2 + \chi_{\rm pull}^2 = \sum_i \sum_\alpha \frac{[N_{i,\alpha}^{\rm th}+ \sum_k C_{i,\alpha k} \xi_k - N_{i,\alpha}^{\rm ex}]^2}{N_{i,\alpha}^{\rm ex}} + \sum_k \xi_k^2,
    \label{eq_chi}
\end{eqnarray}
where $\xi_k$ are the pull variables and $C_{i,\alpha k}$ are the systematic error coefficients corresponding to the pull variables for each flavor.  The pull variables are \textit{flux error}, \textit{zenith error}, \textit{cross-section error}, \textit{systematic error}, and \textit{tilt error}, proportional to the theoretical event rate $N_{i,\alpha}^{\rm th}$m, and are listed in Table~\ref{tab:uncertainty}. The \textit{zenith error} is proportional to the measured angle, i.e., $\cos\theta_m$. We minimize $\chi^2$ by varying the oscillation parameters in the test (th) and obtain $\Delta \chi^2$ relative to the value of $\chi^2$ at the minimum. 
\begin{table}[htbp]
\centering
\caption{\label{tab:uncertainty}Different uncertainties used in our analysis.}
\begin{ruledtabular}
\begin{tabular}{cccc}
Parameters & Percentage\\
\hline
Flux & 20\%  \\
Zenith Angle & 5\%  \\
Cross Section & 10\%  \\                   Systematics & 5\% \\
Tilt & 5\% \\   
\end{tabular}
\end{ruledtabular}
\label{t1}
\end{table}

We used the true and test values of oscillation parameters listed in Table~\ref{tab:NuFit}, following the global NuFit values \cite{Esteban_2024}. We kept $\delta_{\rm CP}=0^\circ$ fixed in both true and test. The total exposure considered is 200 kt$\cdot$yr, which is equivalent to 10 years of runtime with a 20 kton detector module, relevant in the case of a DUNE Phase \uppercase\expandafter{\romannumeral 1}-like setup \cite{DUNE:2020lwj}. 
\begin{table}[htbp]
\caption{\label{tab:NuFit}
The true and test values of the oscillation parameters}
\begin{ruledtabular}
\begin{tabular}{cccc}
Parameters & True (NO)& True (IO)& Test Range\\
\hline
$\theta_{12}$ & $33.68^\circ$ & $33.68^\circ$ & $31.6^\circ$ , $36.0^\circ$ \\
$\theta_{13}$ & $8.56^\circ$ & $8.56^\circ$ & $8.15^\circ$ , $8.9^\circ$ \\
$\theta_{23}$ & $43.3^\circ$ & $43.3^\circ$ & $41.3^\circ$ , $49.9^\circ$ \\
$\Delta m^2_{21}$ & 7.49$\times10^{-5}$ eV$^2$ & 7.49$\times10^{-5}$ eV$^2$ & (6.9, 8.0)$\times10^{-5}$ eV$^2$\\
$\Delta m^2_{31}$ & 2.513$\times10^{-3}$ eV$^2$ & -2.484$\times10^{-3}$ eV$^2$ & $\pm$(2.45, 2.59)$\times10^{-3}$ eV$^2$
\end{tabular}
\end{ruledtabular}
\label{t2}
\end{table}

\section{Sensitivity to the pseudo-Dirac Nature of \texorpdfstring{$\nu_3$}{nu3}}
\label{sec:PD_sensitivity}

\begin{figure}[htbp] 
\centering
\includegraphics[width=0.55\columnwidth]{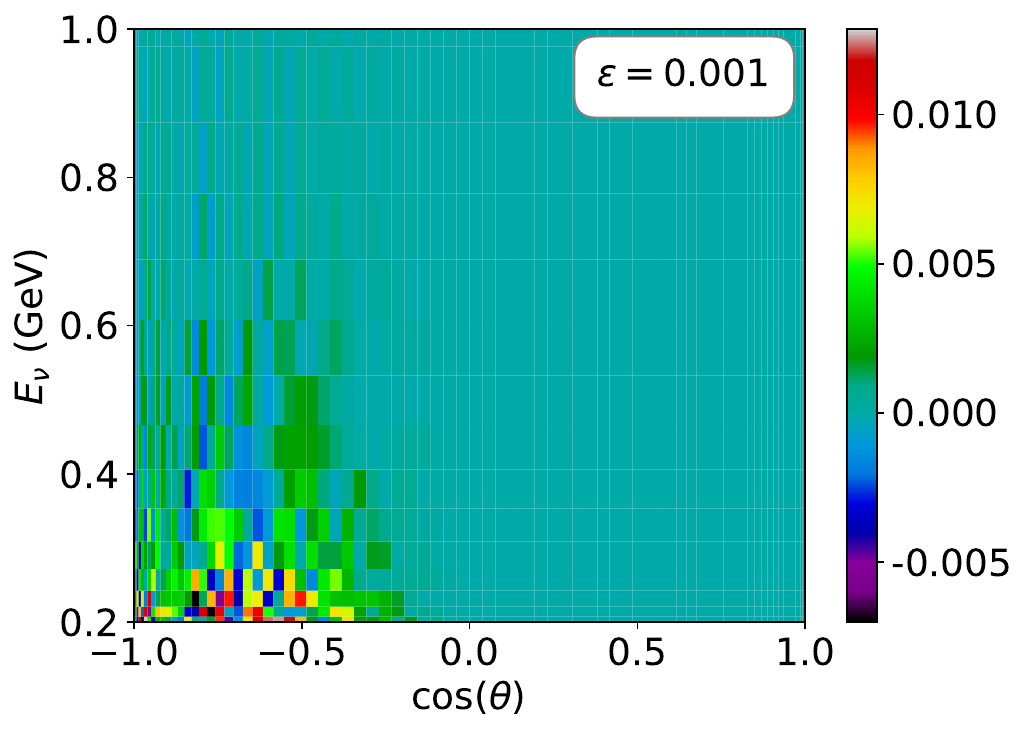}
\includegraphics[width=0.55\columnwidth]{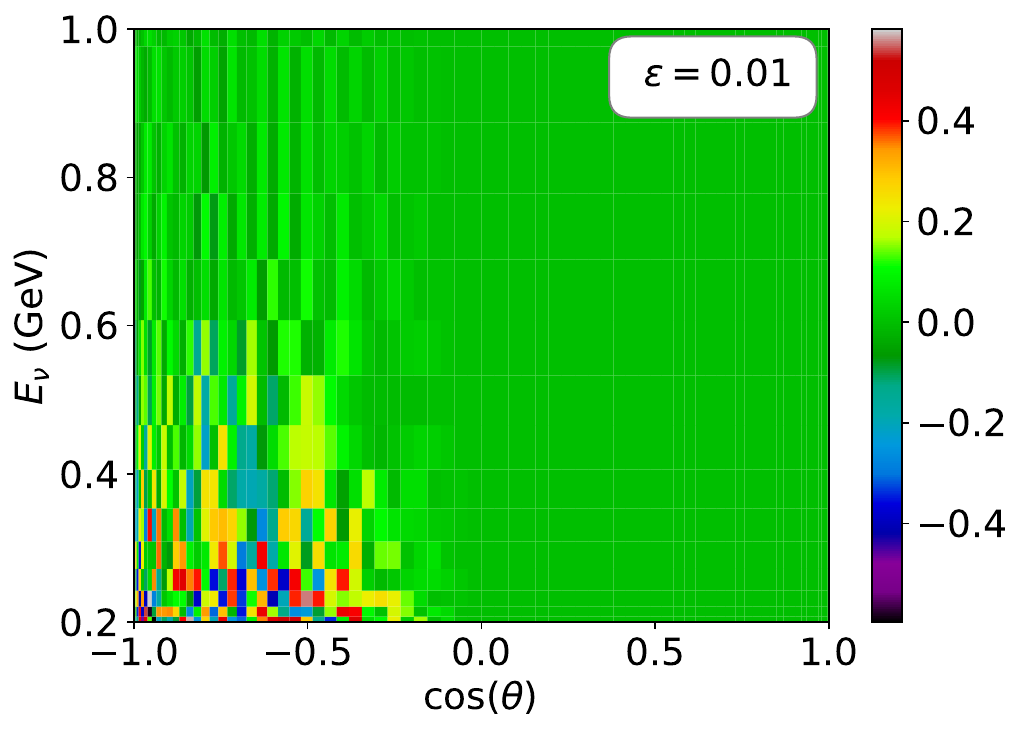}
\includegraphics[width=0.55\columnwidth]{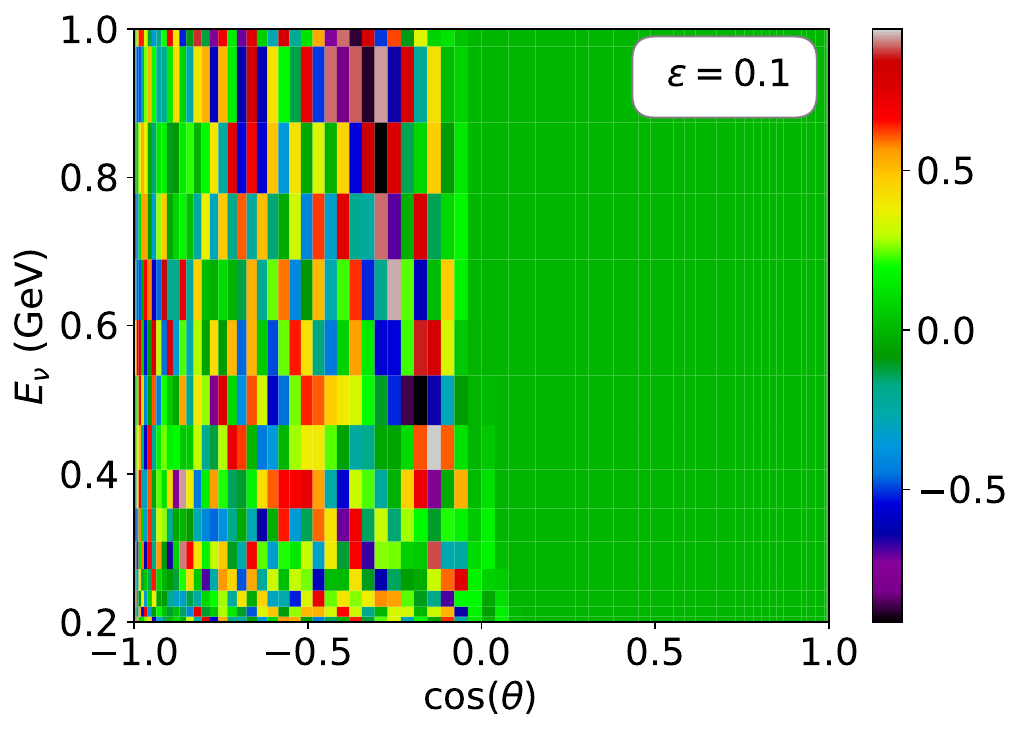}
\caption{The difference of the $\nu_{\mu}$ disappearance probability between the pseudo-Dirac hypothesis and the standard scenario for to $\epsilon=0.001$ (top), $\epsilon=0.01$ (middle), and $\epsilon=0.1$ (bottom), as a function of the neutrino energy $E_{\nu}$ and the neutrino direction $\cos(\theta)$.
  }
\label{prob_PD_OMB}
\end{figure}

Before presenting the results for the sensitivity to the pseudo-Dirac nature of $\nu_3$, we first illustrate in Figure~\ref{prob_PD_OMB} the differences in the oscillograms for the dominant muon-neutrino disappearance probability between the pseudo-Dirac hypothesis and the standard scenario for $\epsilon$ equal to 0.001, 0.01, and 0.1. Non-zero values of the $\epsilon$ parameter lead to distortions at sub-GeV energies and for negative values of $\cos(\theta_z)$. This is not surprising since a large-enough new oscillation phase is required for observable effects and negative $\cos(\theta_z)$ values correspond to longer baselines (and larger matter effects). The effect is seen to be very small -- below a percent -- for $\epsilon=0.001$ but is order one for $\epsilon\gtrsim 0.01$. 

\begin{figure}[htbp] 
\centering
\includegraphics[width=0.42\columnwidth]{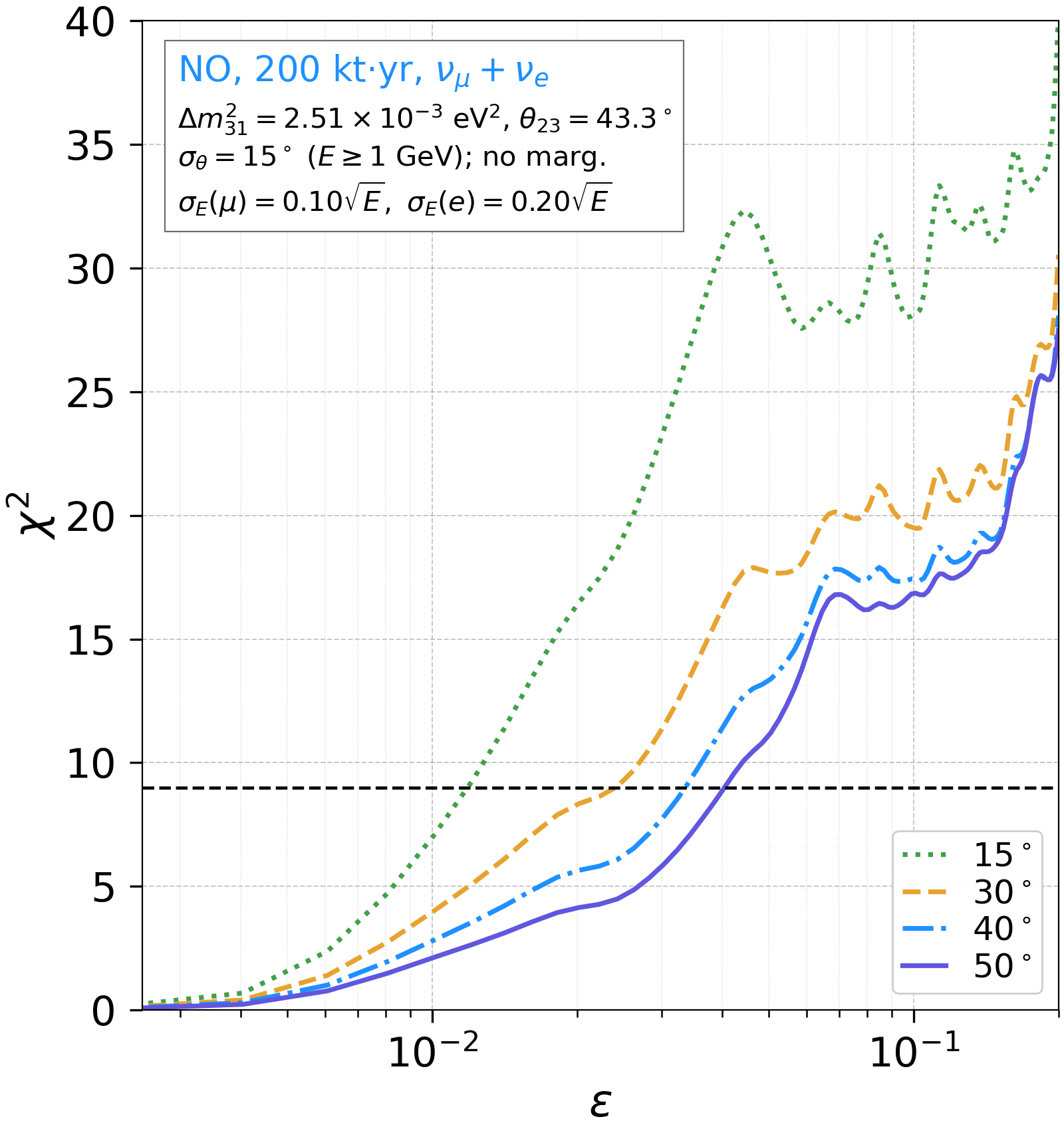}
\includegraphics[width=0.42\columnwidth]{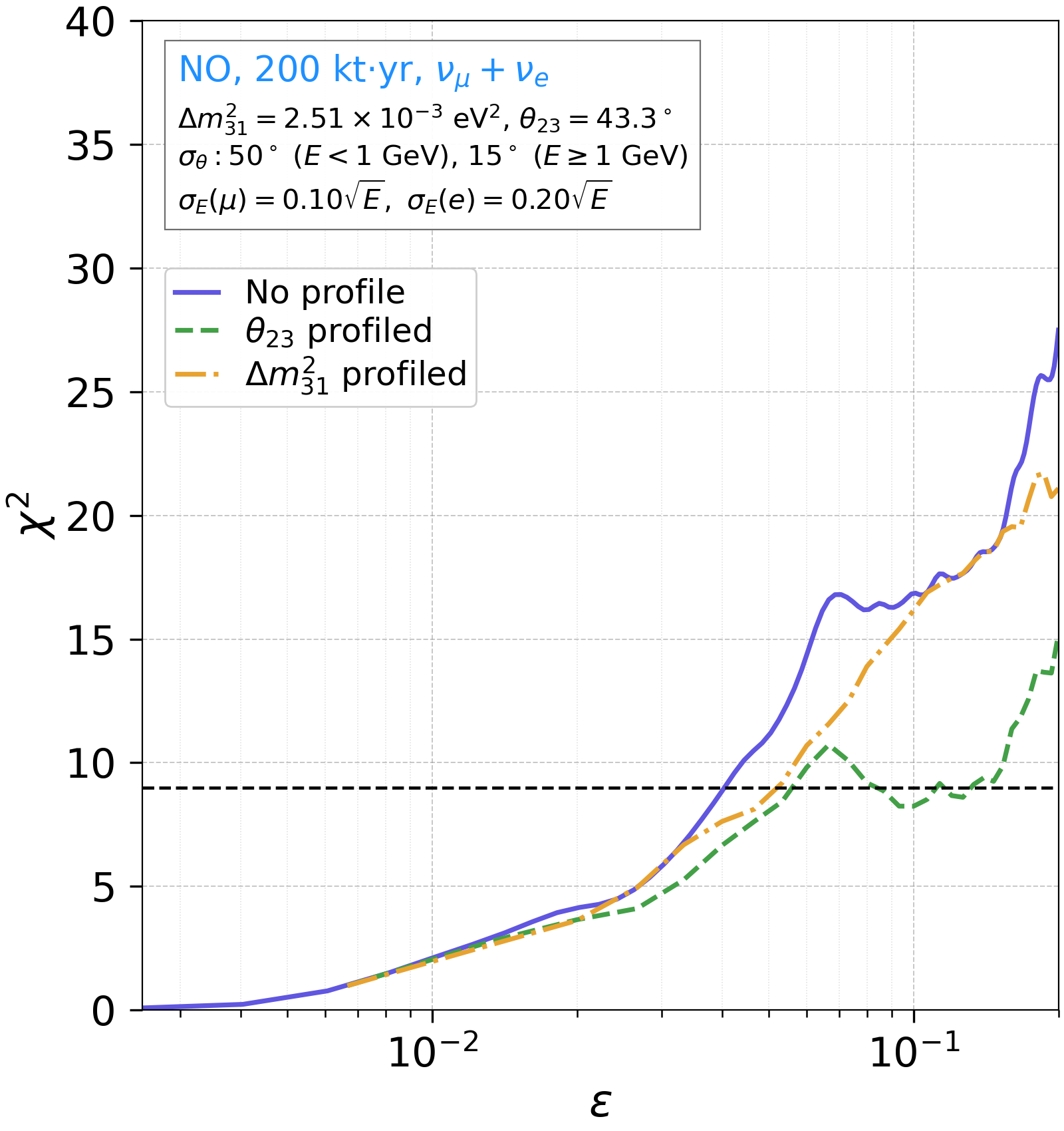}
\caption{Sensitivity to $\epsilon$ of 200~kt$\cdot$yr of atmospheric neutrino data at a DUNE-like detector, assuming NO. On the left is the sensitivity computed for fixed values of all other oscillation parameters for different values of the angular resolution below 1~GeV. 
On the right, we illustrate the effect of profiling over $\theta_{23}$ (green) and $\Delta m^2_{31}$ (gold), relative to the sensitivity obtained for fixed values of all oscillation parameters (blue curve), for a fixed value of the angular resolution below 1~GeV.}
\label{chi-eps_OMB}
\end{figure}

Our estimates for the sensitivity to the pseudo-Dirac nature of $\nu_3$ are depicted in Fig.~\ref{chi-eps_OMB}. We highlight four options for the sub-GeV angular resolution: $15^\circ$,  $30^\circ$, $40^\circ$, and  $50^\circ$. The latter is the most conservative choice in our analysis and was inspired by results made public by the DUNE collaboration~\cite{2026}. We assumed the neutrino mass ordering to be normal (NO); all other assumptions are listed within the panels of Fig.~\ref{chi-eps_OMB}. We present results for the inverted neutrino mass ordering (IO) in Appendix~\ref{app_IO}. We include both the muon-neutrino and electron-neutrino ``data'' in our analyses; the sensitivity, however, is dominated by the muon-neutrino-associated samples.

In the left panel of Fig.~\ref{chi-eps_OMB}, all standard oscillation parameters are fixed to the values listed in Table~ \ref{tab:NuFit}. In this case,  the only free parameter is the value of the dimensionless parameter $\epsilon$ that sets the hypothetical new oscillation frequency. These results indicate that the sub-GeV angular reconstruction significantly impacts the sensitivity. For the most conservative case presented here, a DUNE-like experiment is sensitive to $\epsilon \geq 4.0 \times 10^{-2}$ at the $3\sigma$ level. This corresponds to $\Delta m^2_{33'} \geq 2.0 \times 10^{-4}$ eV$^2$. For the more optimistic case we considered (15$^\circ$ angular smearing below 1 GeV), the 3$\sigma$ sensitivity limit is $\epsilon \geq 1.2 \times 10^{-2}$ which corresponds to $\Delta m^2_{33'} \geq 6.0 \times 10^{-5}$ eV$^2$. 
 For sub-GeV neutrino energies, $\Delta m^2_{33'} \sim 5.0 \times10^{-5}$ eV$^2$ corresponds to an oscillation length of order the Earth's diameter. 

These results are competitive with similar studies in the literature, most notably a recent DUNE beam study~\cite{Anamiati_2018}. This aligns with the fact that atmospheric and beam neutrinos probe, roughly, similar $L/E$ scales. We emphasize, however, that our result indicates a slightly worse relative sensitivity. Essentially, while atmospheric neutrino studies can take advantage of longer baselines and thus can potentially probe relatively small new mass-squared differences, most of the sensitivity lies where the detector angular resolution is the poorest (ie., low neutrino energies). In other words, the access to longer baselines and relatively larger values of $L/E$ is offset by the experiment's ability to determine $L/E$ precisely. We chose to report different instances of sub-GeV angular smearing for this reason.

To probe the robustness of our analysis and to account for the uncertainties present in the current neutrino oscillation global fit values, we also explored the impact of profiling over individual oscillation parameters, taking the current uncertainties quoted in \cite{Esteban_2024} into account. In the right-hand panel of Fig.~\ref{chi-eps_OMB}, we included the effects of profiling, one at a time, over the oscillation parameters that potentially most impact the sensitivity computed here: $\theta_{23}$ and $\Delta m^2_{31}$. Here, the angular resolution for low-energy neutrinos is fixed ($50^\circ$ angular resolution for neutrino energies below 1~GeV). We profile over the so-called atmospheric parameters because they contribute the most to the dominant oscillation amplitude and phase when it comes to the disappearance of atmospheric muon neutrinos.

Our results indicate that profiling over $\theta_{23}$ yields the strongest effect overall, followed by $\Delta m^2_{31}$.
These results are consistent with expectations from a qualitative inspection of the leading-order terms in the oscillation probabilities presented in Appendix~\ref{appA}. 

\begin{figure}[htbp]
    \centering
    \includegraphics[width=0.49\linewidth]{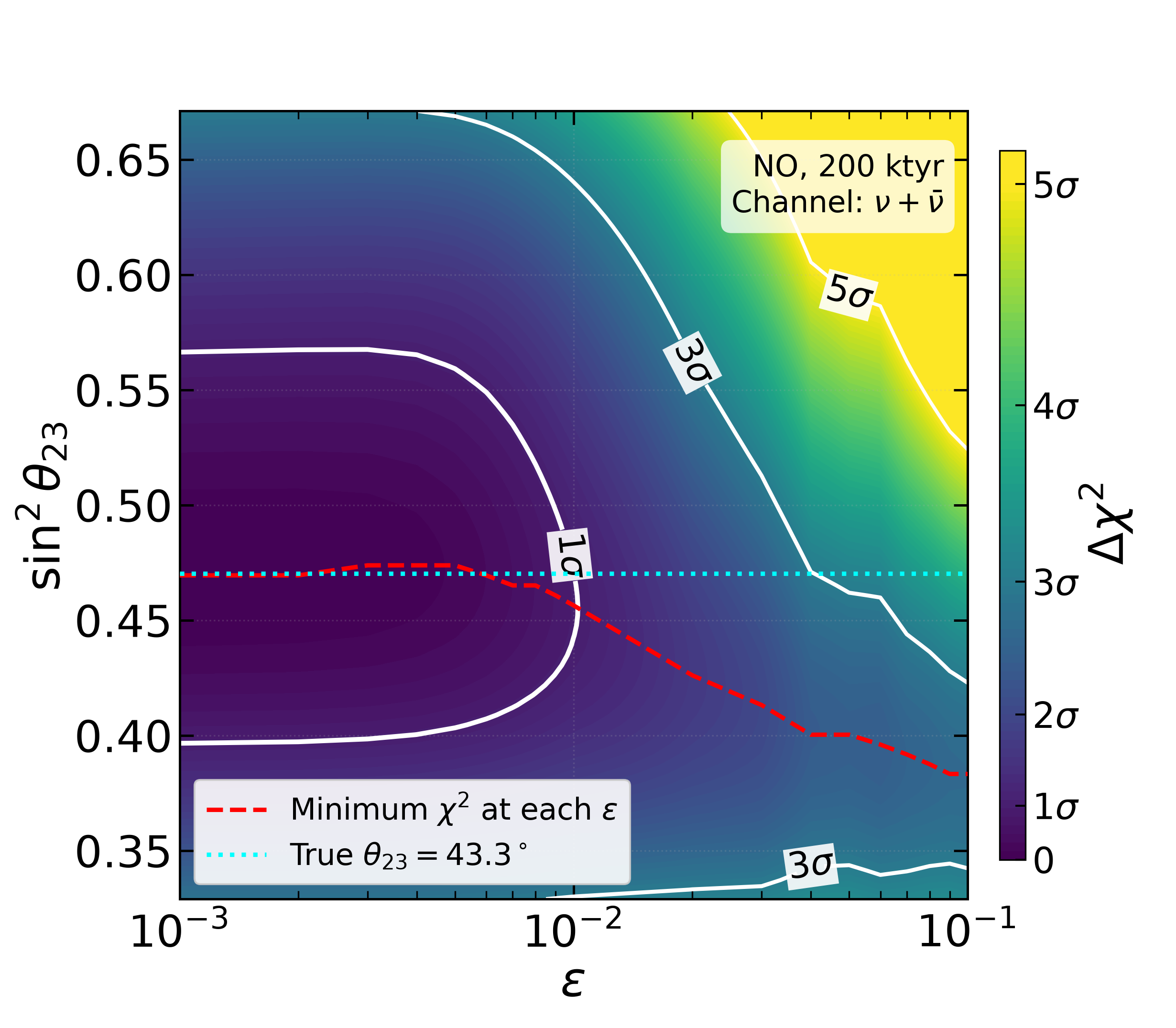}
    \includegraphics[width=0.49\linewidth]{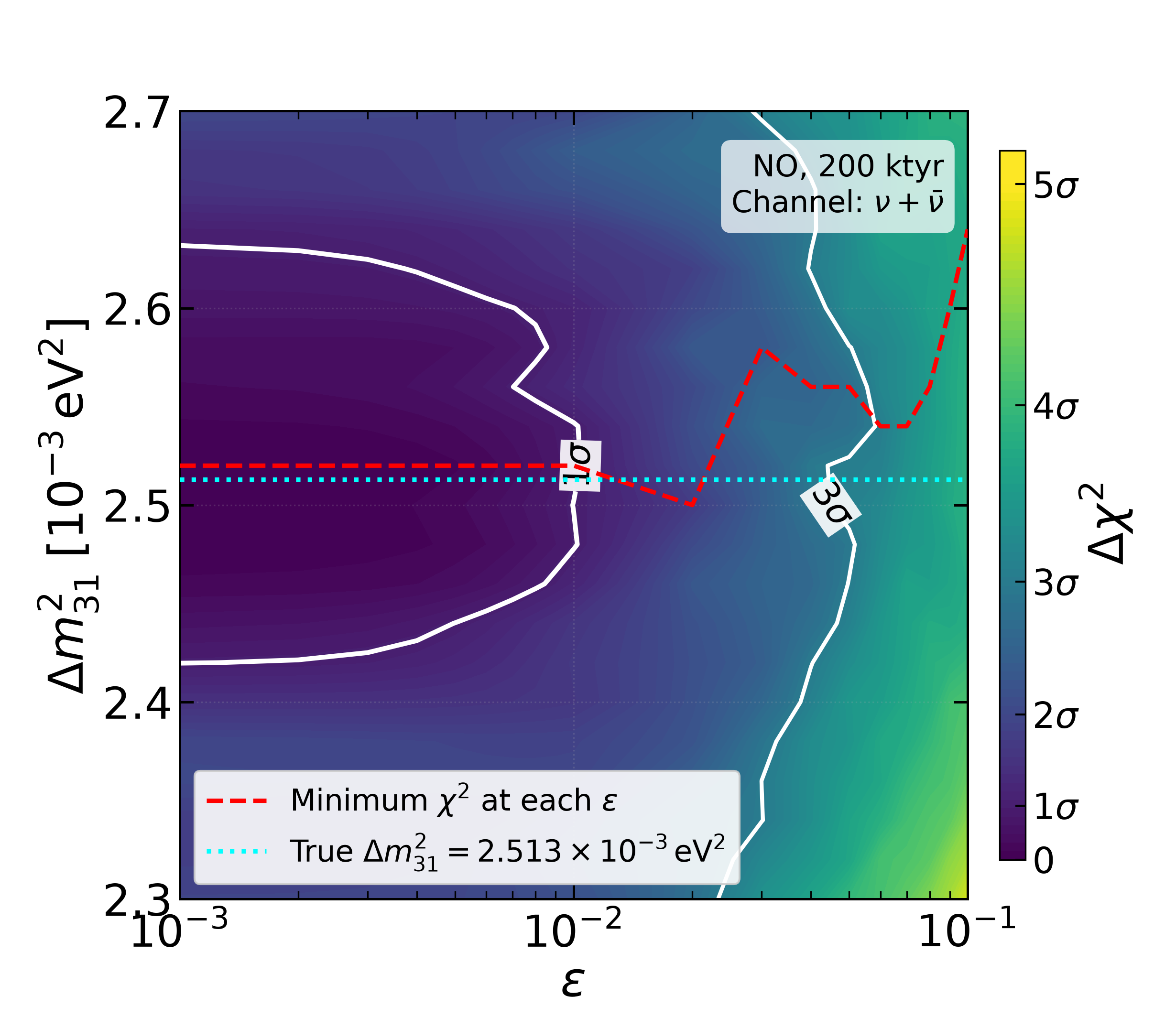}
    \caption{Sensitivity in the $\epsilon\times\sin^2\theta_{23}$ (left), and $\epsilon\times\Delta m^2_{31}$ plane (right) for NO. The red curve shows the position of the minimum value of $\chi^2$ as a function of $\epsilon$. The blue line indicates the true values on the relevant standard oscillation parameter.}
    \label{fig:chi-eps-sth23-dm231-no}
\end{figure}

In a little more detail, terms like $\cos^2\theta_{34}^m\sin2\theta_{23}$, $\sin^2 \theta_{34}^m\sin 2\theta_{23}$, and $\sin^2 2\theta_{34}^m\sin^4\theta_{23}$ in the probabilities indicate a correlation between $\epsilon$ and $\theta_{23}$. This is illustrated in Figure~\ref{fig:chi-eps-sth23-dm231-no}. In the left panel of Fig.~\ref{fig:chi-eps-sth23-dm231-no}, we depict the correlation between $\epsilon$ and $\theta_{23}$. In particular, the red curve depicts the values of the test $\theta_{23}$ at which the minimum $\chi^2$ is obtained for a given value of $\epsilon$. These differ from the true $\theta_{23}=43.3^\circ$ (cyan line) for large-enough values of $\epsilon$; the correlation is clear. Hence, profiling over $\theta_{23}$ yields a relatively lower $\Delta \chi^2$ for higher values of $\epsilon$.  Similarly,  using Eqs.~ (\ref{eq:D31m}, \ref{eq:D32m}) one can see that higher values of $\epsilon$ can be partially compensated by shifting the vale of  $\Delta m^2_{31}$. This is reflected in the red curve in the right-hand panel of Fig.~\ref{fig:chi-eps-sth23-dm231-no}. It indicates that the minimum $\chi^2$ occurs at higher  test-$\Delta m^2_{31}$ values relative to the true value (blue line) as $\epsilon$ increases. Quantitatively, for $\epsilon\lesssim 0.1$, the $\theta_{23}$-related effects are more pronounced.  

\section{Sensitivity to Solar Parameters}
\label{sec:solar}

Since the sensitivity to the pseudo-Dirac mass splitting is close to that of the solar MSW mass-squared difference, it is worthwhile to investigate the sensitivity of the atmospheric neutrinos at a DUNE-like experiment to the solar parameters $\Delta m^2_{21}$ and $\sin^2\theta_{12}$. Similar studies were performed in \cite{Peres:1999yi} in the context of SuperKamiokande data. 

We use as input for the standard $3\nu$-oscillation parameters the values listed in Table~\ref {tab:NuFit}. All other oscillation parameters, apart from the solar under investigation, have been kept fixed in the analysis.

\begin{figure}[htbp]
\centering
\includegraphics[width=0.42\columnwidth]{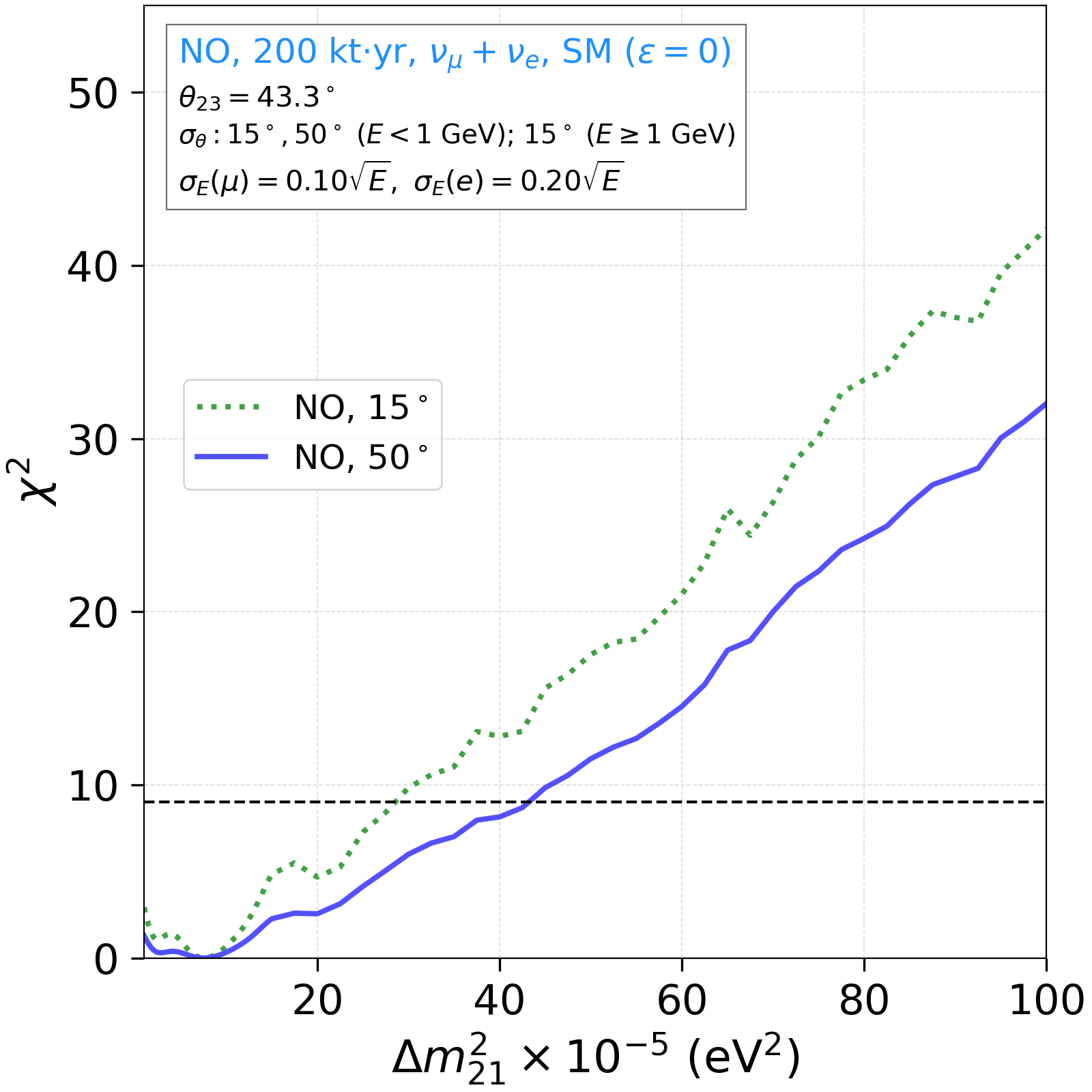}
\includegraphics[width=0.42\columnwidth]{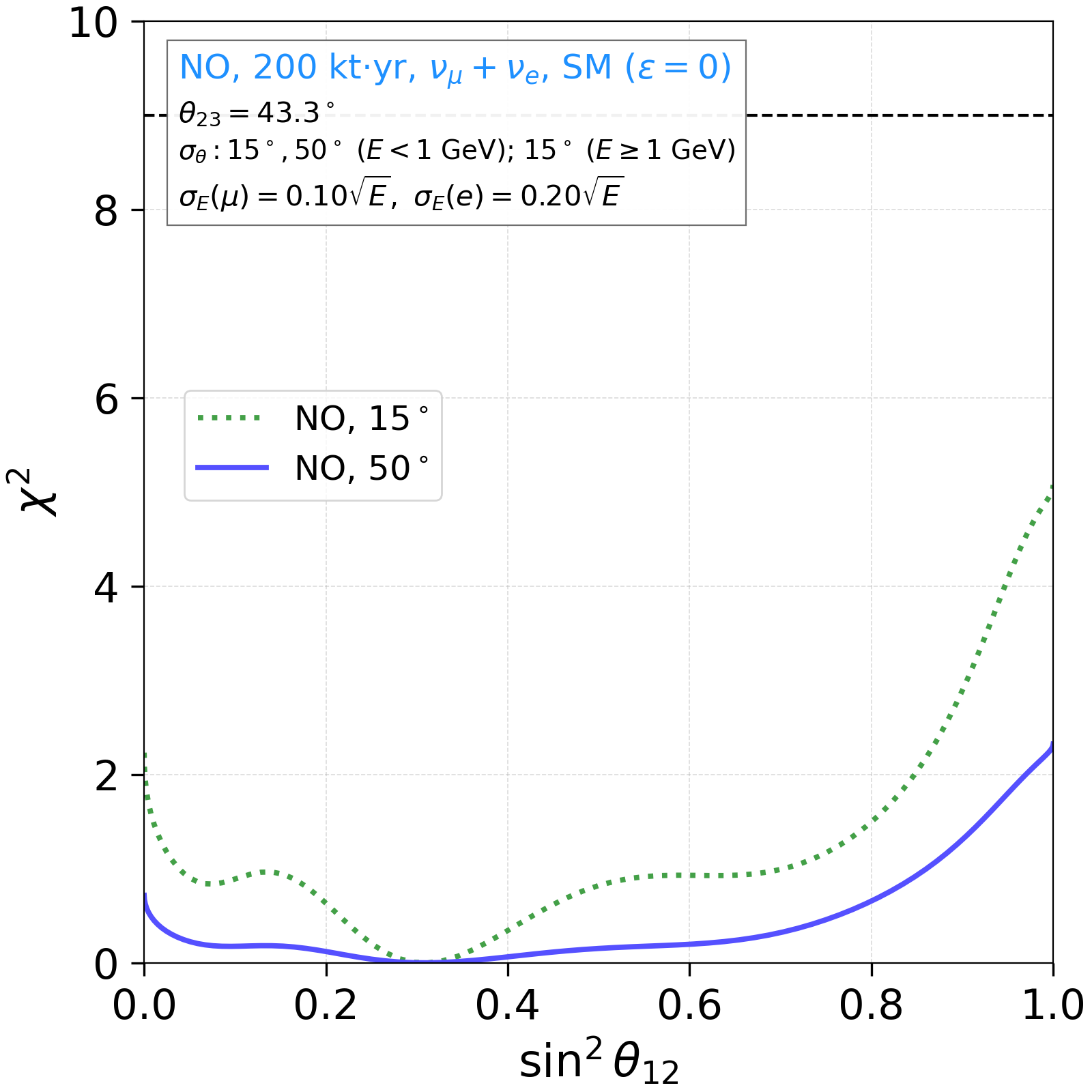}

\caption{Sensitivity to $\Delta m^2_{21}$ (left) and $\sin^2\theta_{12}$ (right) for standard oscillations,  assuming NO and 200 kt$\cdot$yr of data. The dotted and dashed lines correspond to different choices for the angular resolution when $E_\nu< 1$ GeV: $50^\circ$ (most conservative) and $15^\circ$ (most optimistic), respectively.}
\label{chi-dm21_th12-1d-sm_OMB}
\end{figure}

For atmospheric neutrinos, the leading-order dependence on $\theta_{12}$ or $\Delta m^2_{21}$ is subdominant, suggesting weak sensitivity to these parameters. Figure~\ref{chi-dm21_th12-1d-sm_OMB} depicts the sensitivity to $\Delta m^2_{21}$ (left panel) and $\sin^2\theta_{12}$ (right panel). In the left (right) panel, the value of $\theta_{12}$ ($\Delta m^2_{21}$) is kept fixed at the value tabulated in Table~\ref{tab:NuFit}, and we assume NO. There are two curves in each panel indicating the different values of $\vartheta$ that set the angular resolution for $E_{\nu}<1$ GeV: $50^\circ$ and $15^\circ$. As expected, the uncertainty in the direction of low-energy neutrinos has a measurable impact. Atmospheric neutrino data can only set a modest, resolution-dependent limit on $\Delta m^2_{21}$, even when $\Delta m^2_{31}$ is assumed to be perfectly known. The hypothesis $\Delta m^2_{21}=0$ cannot be ruled out at more than $2\sigma$, but values of $\Delta m^2_{21}$ larger than $4\times 10^{-4}$~eV$^2$ can be ruled out at the $3\sigma$ level even when we assume the more pessimistic uncertainty on the direction of low-energy neutrinos. On the other hand, the data can place virtually no bound on $\sin^2\theta_{12}$. 

\begin{figure}[htbp]
\centering
\includegraphics[width=0.48\columnwidth]{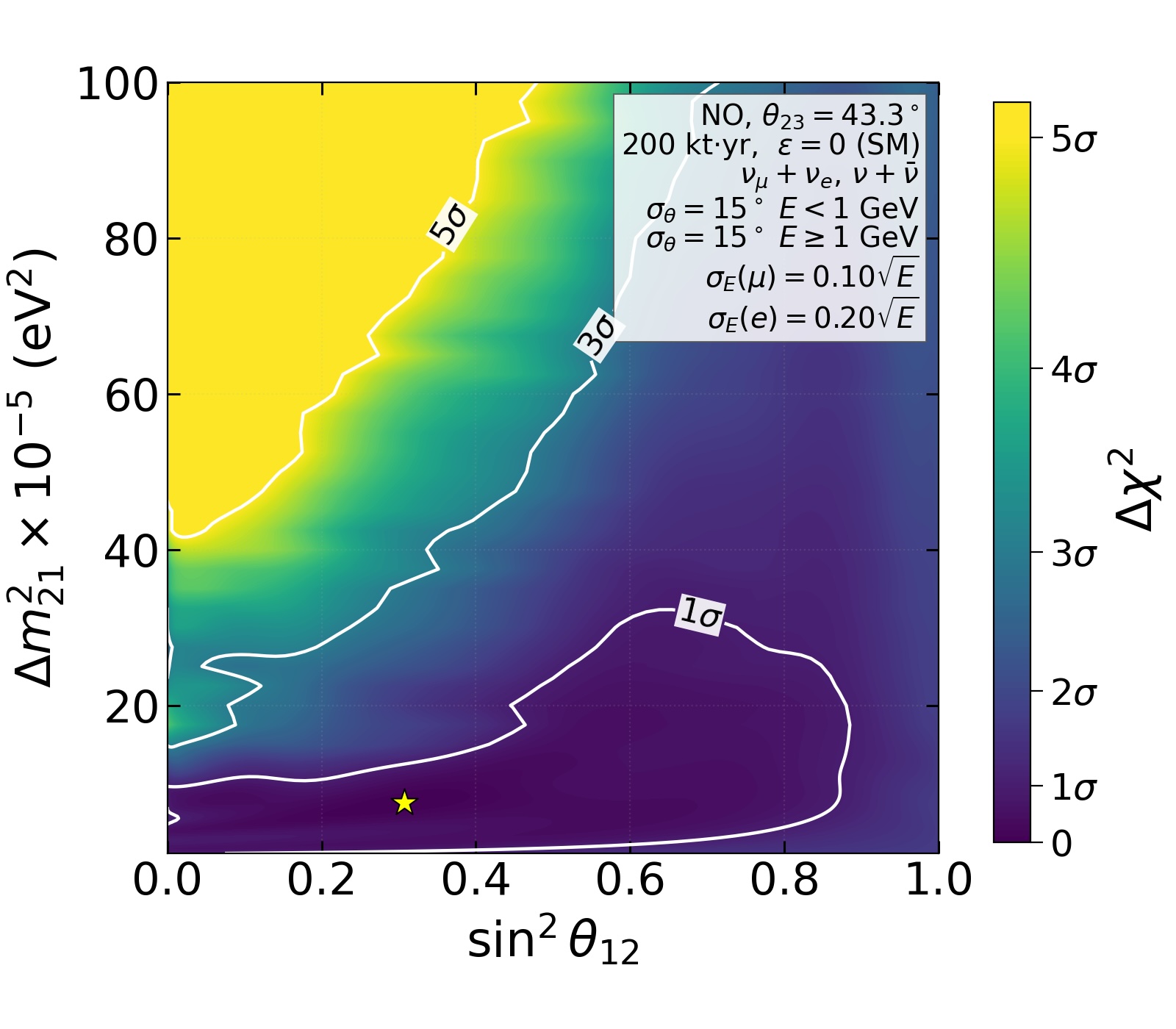}
\includegraphics[width=0.48\columnwidth]{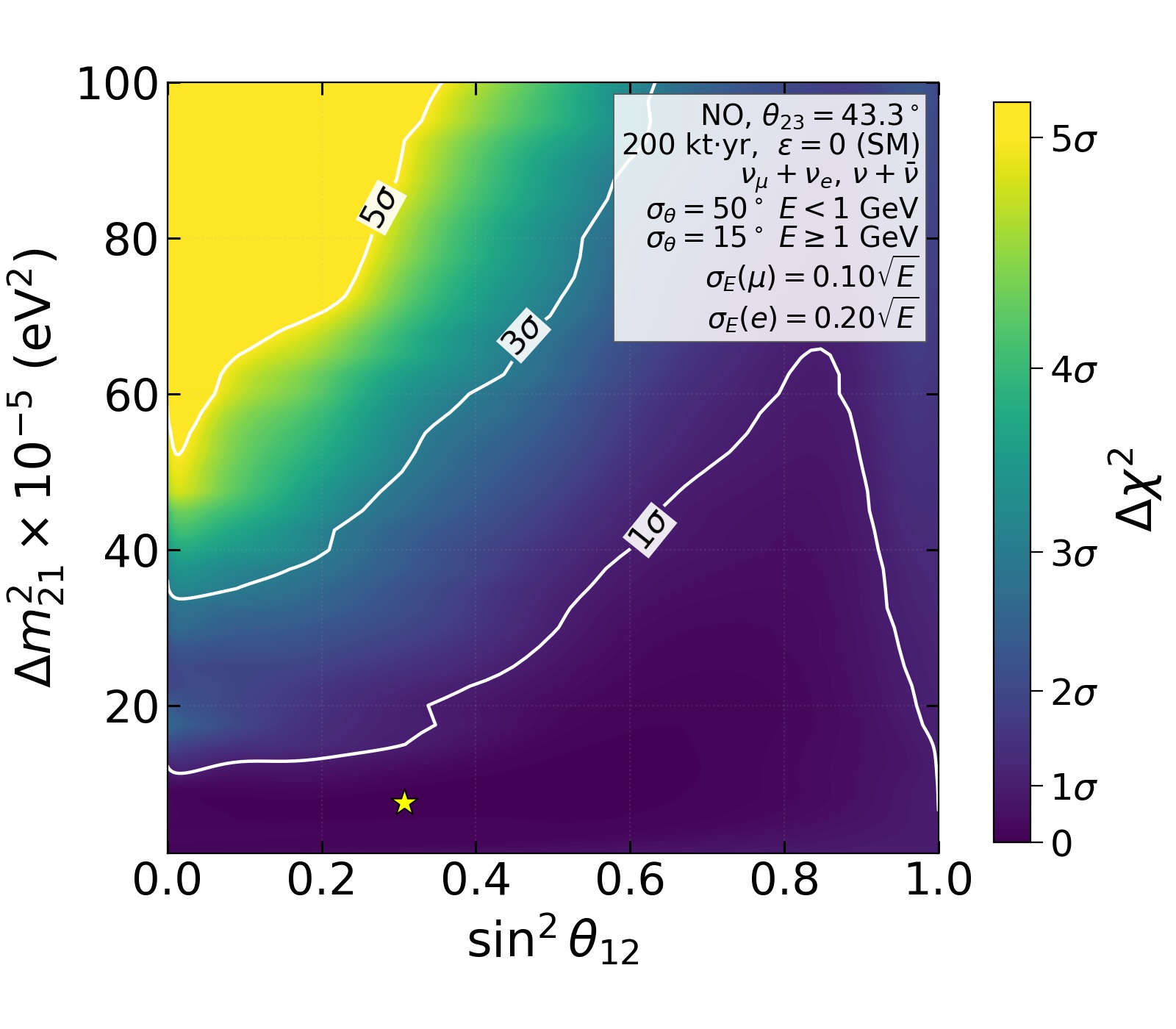}

\caption{Sensitivity to the solar oscillation parameters in the $\Delta m^2_{21}\times\sin^2\theta_{12}$-plane for standard oscillations assuming 200 kt$\cdot$yr of data. We assume NO. The left panel corresponds to an angular resolution for $E_\nu< 1$ GeV of $15^\circ$ (most optimistic), and the right panel corresponds to $50^\circ$ (most conservative).  The star in each panel indicates the best-fit value from NuFit \cite{Esteban_2024}.}
\label{dm21-th12-2d-sm_OMB}
\end{figure}

Fig.~\ref{dm21-th12-2d-sm_OMB} depicts the sensitivity to the solar oscillation parameters in the $\Delta m^2_{12}\times \sin^2\theta_{12}$ plane for two different angular resolutions associated with low-energy atmospheric neutrinos: $15^\circ$ on the left, $50^\circ$ on the right. As before, we have assumed NO. The figure demonstrates strong $\Delta m^2_{21}-\sin^2\theta_{12}$ correlation, as is expected when $L/E$ is small. The sensitivity, even with the most optimistic angular resolution, allows all values of $\sin^2\theta_{12}$ for $\Delta m^2_{21}\lesssim 15\times 10^{-5}$ eV$^2$.

The simulated data is most sensitive to $\nu_{\mu}$ disappearance, so the information on the solar mixing angle comes, in part, from $U_{\mu2}$ (or $U_{\mu1}$; they are related to $U_{\mu3}$ through the unitarity of the mixing matrix). Since we fixed $\delta=0$ throughout, 
\begin{equation}
U_{\mu2} = \cos\theta_{23}\cos\theta_{12} - \sin\theta_{13}\sin\theta_{23}\sin\theta_{12},
\end{equation}
and, numerically,
\begin{equation}
|U_{\mu2}|^2 \simeq 0.5\cos^2\theta_{12} - 0.08
\sin2\theta_{12} + 0.01\sin^2\theta_{12}.
\end{equation}
In particular, $|U_{\mu2}|^2$ decreases as $\sin^2\theta_{12}$ increases for $\sin^2\theta_{12}\lesssim 0.98$.

\section{Conclusions}
\label{sec:conclusions}

We investigated the sensitivity of atmospheric neutrinos at a DUNE-like experiment to the pseudo-Dirac nature of the third neutrino mass eigenstate. We considered the scenario in which the active and sterile components are maximally mixed and the new physics is characterized by a dimensionless parameter, \( \epsilon\), related to the additional mass-squared splitting through \(\Delta m^2_{33'}=2\epsilon\Delta m^2_{31}\). Atmospheric neutrinos are suitable for this search because they cover a broad range of energies (roughly $0.1-100$ GeV) and propagation distances (roughly 20--13000 km) and hence provide access to several orders of \(L/E\) that are not accessible to other conventional terrestrial neutrino sources.

Assuming an exposure of 200~kt$\cdot$yr, we studied the sensitivity to \(\epsilon\), concentrating on the normal neutrino mass ordering. Results for IO can be found in Appendix~\ref{app_IO}. The sensitivity depends strongly on the angular reconstruction of sub-GeV atmospheric-neutrino events. For a sub-GeV angular resolution of \(50^\circ\), the most conservative scenario considered here, if the data are consistent with the standard oscillation scenario, we expect it to translate into a \(3\sigma\) upper bound of 
\[\epsilon < 4.0\times10^{-2},\qquad {\rm or \quad} \Delta m^2_{33'} < 2.0\times10^{-4}\ {\rm eV}^2.\]
Given a  more optimistic angular resolution of \(15^\circ\), future atmospheric data should be able to constrain \[\epsilon < 1.2\times10^{-2},\qquad {\rm or} \qquad \Delta m^2_{33'}< 6\times10^{-5}\ {\rm eV}^2.\]

%
The angular resolution at energies below \(1\ \mathrm{GeV}\) plays a central role in the sensitivity of atmospheric neutrinos to new, long oscillation lengths. 

We further examined the robustness of the sensitivity to the pseudo-Dirac nature of $\nu_3$ against uncertainties in the standard oscillation parameters. Profiling over \(\theta_{23}\) and \(\Delta m^2_{31}\) within the current allowed ranges weakens the sensitivity relative to the case in which all standard parameters are fixed. The effect is more pronounced for  \(\theta_{23}\). A more detailed determination of the pseudo-Dirac sensitivity must therefore account for correlations with the parameters governing the dominant atmospheric oscillation amplitude and frequency. Nevertheless, the atmospheric-neutrino reach remains comparable to that of related terrestrial searches and provides complementary information.

We also explored the sensitivity of atmospheric neutrinos in a DUNE-like detector to the solar parameters \(\Delta m^2_{21}\) and \(\sin^2\theta_{12}\). This question is similar to that of the pseudo-Dirac nature of $\nu_3$, i.e., it is also related to the sensitivity to new long oscillation lengths. We find that the sensitivity to the solar oscillation parameters is limited. This is consistent with the bounds on $\Delta m^2_{33'}$ summarized above, which not coincidentally, are of order $\Delta m^2_{21}$.

Our analysis highlights the potential and limitations of atmospheric neutrinos as complementary probes of the pseudo-Dirac nature of light neutrinos. It underscores the importance of the angular reconstruction of sub-GeV neutrinos, control of systematic uncertainties, and the treatment of correlations among the known oscillation parameters as central ingredients in determining the final experimental reach. 
 
We anticipate that atmospheric neutrinos collected by the Hyper-Kamiokande experiment \cite{Hyper-Kamiokande:2018ofw} should have similar sensitivity to the pseudo-Dirac nature of $\nu_3$. Ultimately, the combination of atmospheric-neutrino observations with accelerator, reactor, solar, and astrophysical measurements will provide coverage across several orders of magnitude in pseudo-Dirac mass splittings.

{\it{Note added}}: While some of the authors are members of the DUNE collaboration, this work is not officially associated with DUNE; only publicly available detector specifications were used to simulate and analyze hypothetical data in a DUNE-like detector. 

\begin{acknowledgments}
SG acknowledges Northwestern University, where the work was initiated, for hospitality and the Fulbright-Nehru Academic and Professional Excellence fellowship for funding her visit to the US.
SP acknowledges the funding from ANID-Chile under FONDECYT Postdoctorado No. 3250376. The work of OMB and AdG was supported in part by the US Department of Energy (DOE) grants \#de-sc0010143 and \#de-sc0010813 and in part by the National Science Foundation under Grant Nos.~PHY-1630782 and PHY-1748958.

\end{acknowledgments}

\appendix

\section{Approximate Analytic Expressions for the Oscillation Probabilities \label{appA}}

We can write the total Hamiltonian that governs the 3+1 system of interest as the sum of two Hamiltonians,
    \begin{equation}
        \mathcal{H}_{3+1}=\frac{1}{2E}\left(\mathcal{H}^{(0)} +\mathcal{H}^{(1)}\right).
    \end{equation}
$\mathcal{H}^{(0)}$ is zeroth-order Hamiltonian proportional to $\Delta m^2_{31}$ and $\mathcal{H}^{(1)}$ is the perturbative Hamiltonian containing the small terms proportional to $\epsilon$ and $\Delta m^2_{21}$. We use perturbation theory, considering $\Delta m^2 _ {21} $ (in units of $\Delta m^2_{31}$) and $ \epsilon$ as small parameters, to calculate the eigenvalues of the Hamiltonian and the oscillation frequencies in matter. Throughout this appendix, we assume the matter potential $A$ is constant. 
The eigenvalues of $\mathcal{H}_{3+1}/2E$ are,
\begin{align}
\lambda_1^{m} &=
\frac{1}{2}\left(\Delta m^2_{21}+A c_{13}^{2}-\Delta_{21}^{m}\right)\label{eq:l1m},
\\[2mm]
\lambda_2^{m}&=
\frac{1}{2}\left(\Delta m^2_{21}+A c_{13}^{2}+\Delta_{21}^{m}\right)\label{eq:l2m},
\\[2mm]
\lambda_3^{m}&=
\Delta m^2_{31}+\frac{1}{4} \left(2A-A\cos 2\theta_{13}\right) -\frac{1}{2}\Delta_{43}^{m}\label{eq:l3m},
\\[2mm]
\lambda_4^{m}&=
\Delta m^2_{31}+\frac{1}{4}\left(2A-A\cos 2\theta_{13}\right)+\frac{1}{2}\Delta_{43}^{m}\label{eq:l4m},
\end{align}
where the modified mass-squared differences are,
\begin{align}
\Delta_{21}^{m} & =
\sqrt{\left(\Delta m^2_{21}\cos 2\theta_{12} - A c_{13}^{2}\right)^{2}
+
\left( \Delta m^2_{21}\sin2\theta_{12} \right)^{2}}\label{eq:D21m},
\\[2mm]
\Delta_{43}^{m} &  =
\sqrt{A^{2}\cos^{2}2\theta_{13} +
\left( 4\Delta m^2_{31}\epsilon \right)^{2}
} \equiv \sqrt{A^{2}\cos^{2}2\theta_{13} +
(2\Delta m^2_{33'})^{2}
}\label{eq:D43m},
\\[2mm]
\Delta_{31}^{m} & = 
\Delta_{31}
+\frac{1}{4}
\left(
2A-A\cos2\theta_{13}
\right)
-\frac{1}{2}\Delta_{43}^{m}
-\frac{1}{2}
\left(
\Delta_{21}+Ac_{13}^{2}-\Delta_{21}^{m}
\right)\label{eq:D31m}\\[2mm]
\Delta_{32}^{m} & = 
\Delta_{31}
+\frac{1}{4}
\left(
2A-A\cos2\theta_{13}
\right)
-\frac{1}{2}\Delta_{43}^{m}
-\frac{1}{2}
\left(
\Delta_{21} + Ac_{13}^{2} + \Delta_{21}^{m} \right)\label{eq:D32m}.
\end{align}

We can write the mixing matrix in matter, obtained from the eigenvectors of the Hamiltonian, as
\begin{equation}\label{eq:Upd}
U_{\rm pd} = R_{23}.\begin{pmatrix}
c_{12}^{m} c_{13}^{m}
& s_{12}^{m} c_{13}^{m}
& s_{13}^{m}c_{34}^{m}
& s_{13}^{m}s_{34}^{m}
\\[3mm]
-s_{12}^{m} & c_{12}^{m} & 0 & 0
\\[3mm]
-c_{12}^{m}s_{13}^{m}
& -s_{12}^{m}s_{13}^{m}
& c_{13}^{m}c_{34}^{m}
& c_{13}^{m}s_{34}^{m}
\\[3mm]
0 & 0 & -s_{34}^{m} & c_{34}^{m}
\end{pmatrix} \equiv R_{23}(\theta_{23}).R_{12}^{m}(\theta_{12}^m).R_{13}^{m}(\theta_{13}^m).R_{34}^{m}(\theta_{34}^m).
\end{equation}
Along with the ``matter'' mixing angles $\theta_{12}^m$ and $\theta_{13}^m$, there is also an effective mixing angle $\theta_{34}^m$. It turns out (cf. Eq.~(\ref{eq:th34m}))  that $\theta_{34}^m$ is very small  when $\Delta m^2_{31}\epsilon<<A\cos 2\theta_{13}$. Therefore, terms in the oscillation probabilities proportional to $\sin^2\theta_{34}^m$ will be sub-leading. Using perturbation theory, the modified mixing angles satisfy
\begin{eqnarray}\label{eq:th12m}
    \sin\theta_{13}^m = \sin\theta_{13}\left[ 1 + \frac{A \cos^2{\theta_{13}}}{\Delta m^2_{31}}\right], \,
    \cos\theta_{13}^m =\cos\theta_{13}\left[ 1 - \frac{A \sin^2{\theta_{13}}}{\Delta m^2_{31}}\right].
\end{eqnarray}
\begin{align}\label{eq:th12mb}
\sin 2\theta_{12}^{m} & =
\frac{\Delta m^2_{21}\sin 2\theta_{12}}{\Delta_{21}^{m}},
\,
\cos 2\theta_{12}^{m} =
\frac{\Delta m^2_{21}\cos 2\theta_{12} - A \cos^2\theta_{13}}{\Delta_{21}^{m}}.
\end{align}
\begin{align}\label{eq:th34m}
\sin 2\theta_{34}^{m} &= \frac{4\Delta m^2_{31}\epsilon}{\Delta_{43}^{m}} \equiv \frac{2\Delta m^2_{33'}}{\Delta_{43}^{m}},
\,
\cos 2\theta_{34}^{m} =
\frac{A\cos 2\theta_{13}}{\Delta_{43}^{m}}.
\end{align}

The matter equivalent version of $\Delta m^2_{21}$  is modified relative to its vacuum value by a function of the matter potential $A$ and the mixing angles $\theta_{13}$ and $\theta_{12}$. On the other hand, the matter equivalent to the atmospheric mass-squared difference is a function of both the matter potential and the size of the pseudo-Dirac splitting, proportional to $\epsilon$. A modified mass-squared difference $\Delta_{43}^m$  also arises in matter and is a function of $A$ and $\epsilon$. However, it is a stronger function of $A$, as seen in Eq.~(\ref{eq:D43m}).

To express the oscillation probabilities in constant matter, we define
\begin{eqnarray}
P^0 &=&
s_{13}^{m} \cos\left(2\theta_{12}^{m}\right)
\sin\left(2\theta_{23}\right)
+
\sin\left(2\theta_{12}^{m}\right)
\Big(
c_{23}^{2} - \left(s_{13}^{m}\right)^{2}s_{23}^{2}
\Big),\label{eq:P0}
\\
P_{\mu\mu}^1&=& \Big[s_{12}^{m}c_{13}^{m}
\sin\left(2\theta_{23}\right)
+
c_{12}^{m} s_{23}^{2}
\sin\left(2\theta_{13}^{m}\right) \Big]^2,
\label{eq:Pmm1}\\
P_{\mu\mu}^2&=&\left(c_{13}^{m}\right)^{2}
\sin^{2}\left(2\theta_{23}\right) +
s_{23}^{4}\sin^{2}\left(2\theta_{13}^{m}\right),
\label{eq:Pmm2}\\
P_{\mu e}^1&=& \left(c_{13}^{m}\right)^{2}s_{13}^{m}
\sin\left(2\theta_{23}\right)
\sin\left(2\theta_{12}^{m}\right)
+
\left(c_{12}^{m}\right)^{2} s_{23}^{2}
\sin^{2}\left(2\theta_{13}^{m}\right).
\label{eq:Pme1}
\end{eqnarray}
The appearance and disappearance probabilities in matter in the presence of a pseudo-Dirac-related $\Delta m^2_{33'}$ are
\begin{align}\label{eq:Pme-pd}
P_{\mu e}
={}&
\left(c_{13}^{m}\right)^{2}\sin\left(2\theta_{12}^{m}\right)
P^0 \sin^{2}\Phi^m_{21}
+
\left(c_{34}^{m}\right)^{2}
\Bigg[
P_{\mu e}^1
\left( \sin^2 \Phi^m_{31} - \sin^2 \Phi^m_{32} \right) 
+ s^2_{23} \sin^2\left(2\theta_{13}^{m}\right)\sin^{2}\Phi^m_{32}
\Bigg]
\nonumber\\&
+\left(s_{34}^{m}\right)^{2}
\Bigg[
P_{\mu e}^1
\left(\sin^{2}\Phi^m_{41} - \sin^{2}\Phi^m_{42}\right)
+
\sin^{2}\left(2\theta_{13}^{m}\right) s_{23}^{2}
\Big(
\sin^{2}\Phi^m_{42} - \left(c_{34}^{m}\right)^2
\sin^{2}\Phi^m_{43} \Big)
\Bigg],\\
\label{eq:Pee-pd}
P_{ee}
=1-
&
\left(c_{13}^{m}\right)^{4} \sin^2\left(2\theta_{12}^{m} \right)\sin^{2}\Phi^m_{21}
+
\left(c_{34}^{m}\right)^{2}
\Bigg[
\left(c_{12}^{m}\right)^{2} \sin^2\left(2\theta_{13}^{m}\right)\sin^{2}\Phi^m_{31}
+
\left(s_{12}^{m}\right)^{2}
\sin^2 \left(2\theta_{13}^{m}\right)\sin^{2}\Phi^m_{32}
\Bigg]
\nonumber\\
&+
\left(s_{34}^{m}\right)^{2}
\Bigg[
\left(c_{12}^{m}\right)^{2} \sin^2\left(2\theta_{13}^{m}\right)\sin^{2}\Phi^m_{41}
+
\left(s_{12}^{m}\right)^{2}
\sin^2 \left(2\theta_{13}^{m}\right)\sin^{2}\Phi^m_{42}
\Bigg]
+
\left(s_{13}^{m}\right)^{4} \sin^2\left(2\theta_{34}^{m}\right)\sin^{2}\Phi^m_{43},\\
\label{eq:Pmm-pd}
P_{\mu\mu}
={}&
1-(P^0)^{2}
\sin^{2}\Phi^m_{21}
-\left(c_{34}^{m}\right)^{2}
\Bigg[
P_{\mu\mu}^1\left(\sin^{2}\Phi^m_{31} - \sin^{2}\Phi^m_{32}\right)
+
P_{\mu\mu}^2 \sin^{2}\Phi^m_{32}
\Bigg]
\nonumber\\
&
-\left(s_{34}^{m}\right)^{2}
\Bigg[
P_{\mu\mu}^1 \left(\sin^{2}\Phi^m_{41} - \sin^{2}\Phi^m_{42} \right)
+ P_{\mu\mu}^2 \sin^{2}\Phi^m_{42}
\Bigg]
-\left(c_{13}^{m}\right)^{4}
s_{23}^{4}
\sin^{2}\left(2\theta_{34}^{m}\right)
\sin^{2}\Phi^m_{43}.
\end{align}
where $\Phi^m_{ij}\equiv \Delta_{ij}^{m}L/4E$.

A few points related to the expressions above are worthy of note: 
\begin{itemize}
   \item The new oscillation frequencies $\Phi^m_{41},\Phi^m_{42},\Phi^m_{43}$ terms are subleading due to the smallness of  $\sin^2\theta_{34}^m$,

\item The presence of terms with both $\theta_{34}^m$ and $\theta_{23}$ in the amplitude, e.g., $\sin^2 2\theta_{34}^m$ together with $\sin^4 \theta_{23}$ in $P_{\mu \mu}$, translate into a correlation between $\epsilon$ and $\theta_{23}$; an increase in $\epsilon$ (and hence $\sin^2 2\theta_{34}^m$)  can be compensated by a decrease in $\theta_{23}$. 
   
   \item 
    There is a correlation between $\Delta m^2_{31}$ and $\Delta m^2_{33'}$. These can be in the expressions for $\Delta_{31}^m$, and $\Delta_{32}^m$ (using $\Delta^m_{43}$); 
    larger values of $\Delta m^2_{31}$ can be compensated by an increase in $\epsilon$ and hence of $\Delta_{43}^m$. 
   
\end{itemize}

For completeness, we list the 
the expressions for the oscillation probabilities in the standard oscillation scenario -- three active flavors -- under the same conditions:
\begin{align}\label{eq:Pab-pd}
P_{\mu e}
&=
\left(c_{13}^{m}\right)^{2}\sin\left(2\theta_{12}^{m}\right)
P^0 \sin^{2}\Phi_{21}^m
+
\Bigg[
P_{\mu e}^1
\left( \sin^2 \Phi_{31} - \sin^2 \Phi_{32} \right) 
+ s^2_{23} \sin^2\left( 2\theta_{13}^{m} \right) \sin^{2}\Phi_{32}
\Bigg],\\
P_{ee}
&=1-
\left(c_{13}^{m}\right)^{4} \sin^2\left(2\theta_{12}^{m} \right)\sin^{2}\Phi_{21}^m
+ \sin^2 \left(2\theta_{13}^{m}\right)
\Bigg[
\left(c_{12}^{m}\right)^{2} \sin^{2}\Phi_{31}
+
\left(s_{12}^{m}\right)^{2} \sin^{2}\Phi_{32}
\Bigg],\\
P_{\mu\mu}
&=
1-(P^0)^{2}
\sin^{2}\Phi_{21}^m-
\Bigg[P_{\mu\mu}^1\left(\sin^{2}\Phi_{31} - \sin^{2}\Phi_{32}\right)
+ P_{\mu\mu}^2 \sin^{2}\Phi_{32} \Bigg],
\end{align}
where $\Phi_{31}$ and $\Phi_{32}$ are
\begin{align}
    \Phi_{31}&=\frac{L}{4E}\left( \Delta_{31}+\frac{1}{4} \left(2A-A\cos 2\theta_{13}\right) - \frac{1}{2}\left(\Delta_{21}+A c_{13}^{2}-\Delta_{21}^{m}\right) \right),\\
    \Phi_{32}&=\frac{L}{4E}\left( \Delta_{31}+\frac{1}{4} \left(2A-A\cos 2\theta_{13}\right) - \frac{1}{2}\left(\Delta_{21}+A c_{13}^{2}+\Delta_{21}^{m}\right) \right).
\end{align}

\section{Results for the Inverted Ordering}
\label{app_IO}

Here we collect the results obtained assuming that the neutrino mass ordering is inverted. Note that, in this case, the connection between $\Delta m^2_{33'}$ and $|\Delta m^2_{31}|$ is not especially natural since, unless all neutrino masses are degenerate, $m_3^2\ll |\Delta m^2_{31}|$.

\begin{figure}[htbp] 
\centering
\includegraphics[width=0.42\columnwidth]{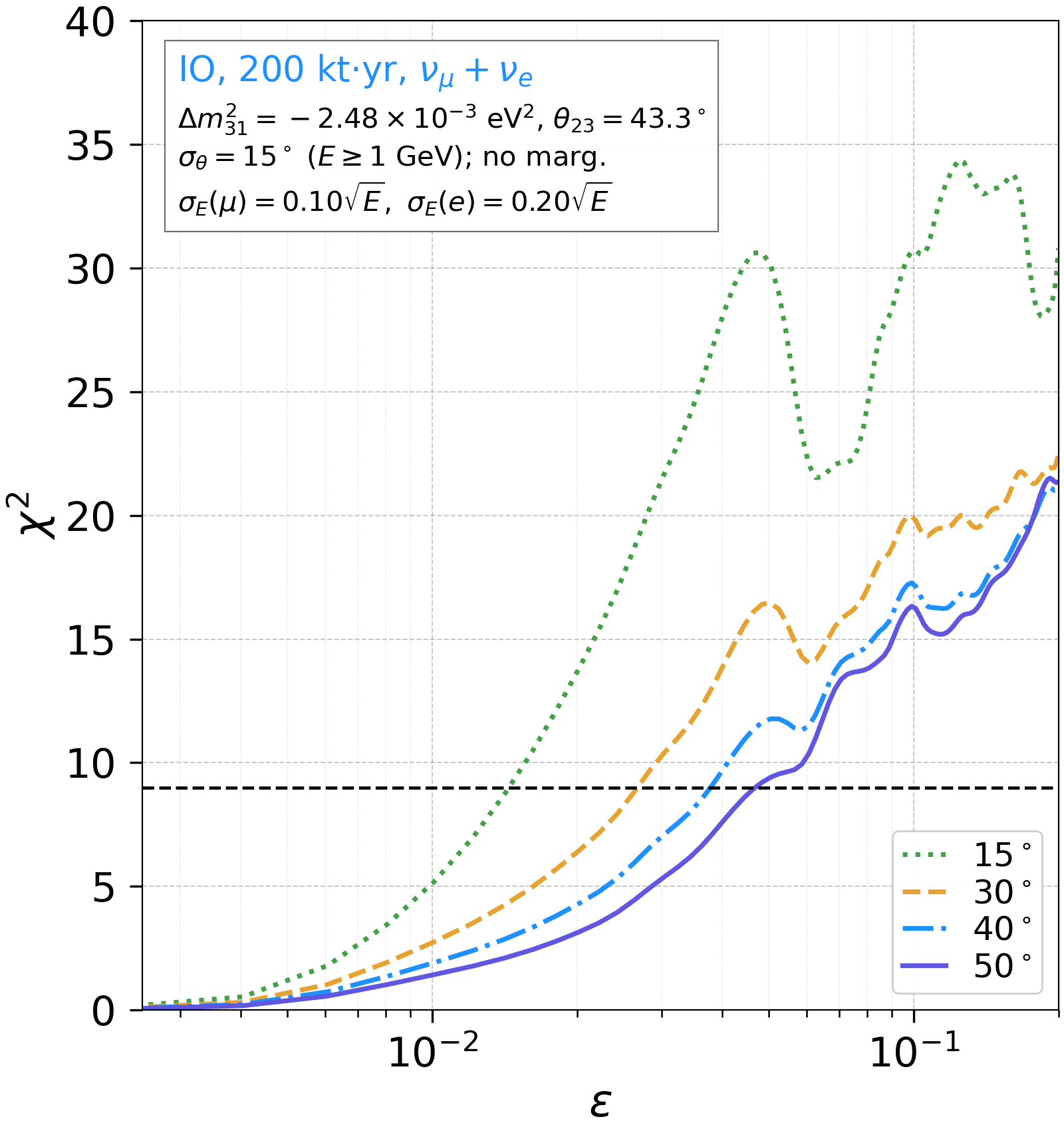}
\includegraphics[width=0.42\columnwidth]{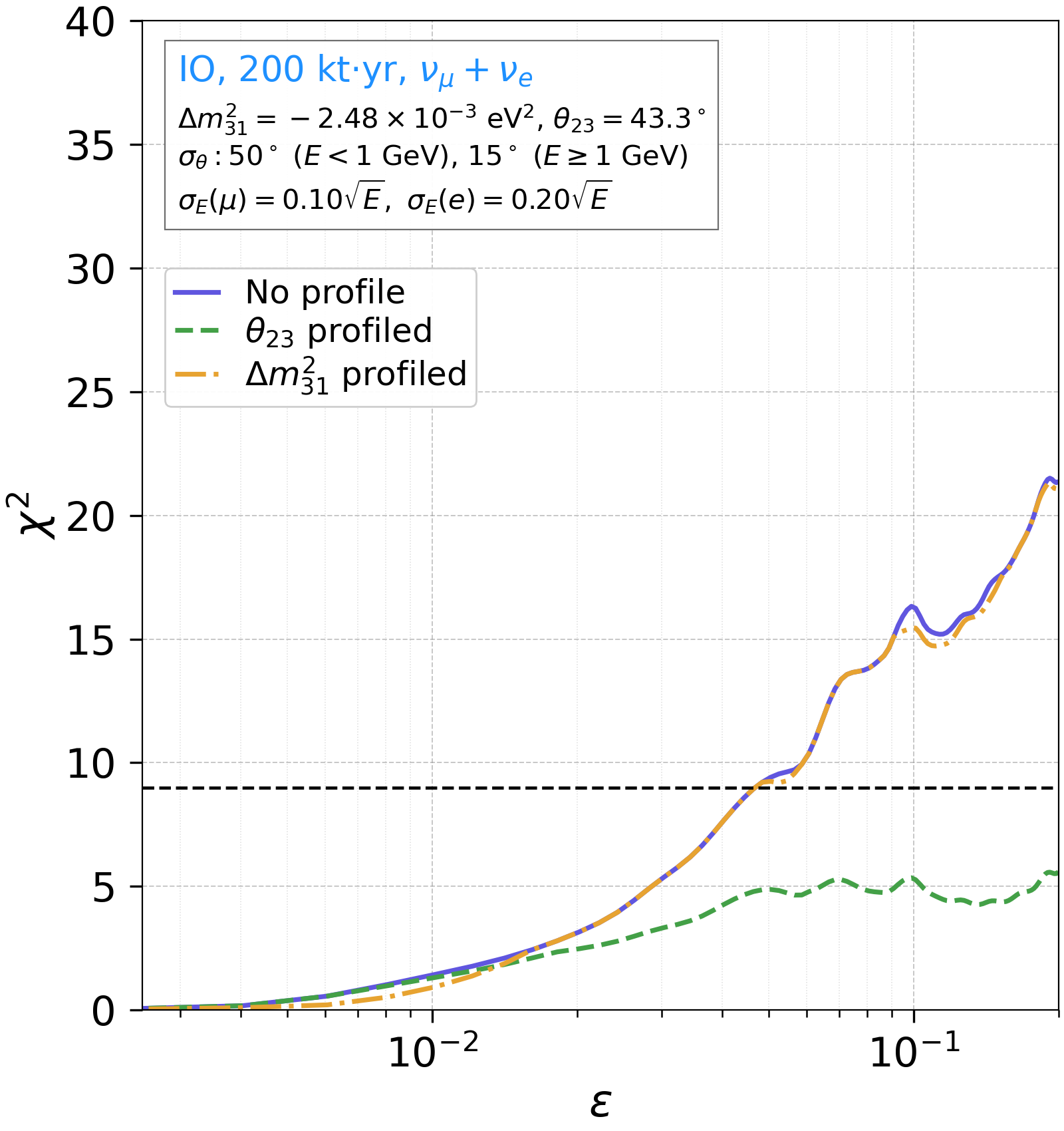}
\caption{Sensitivity to $\epsilon$ of 200~kt$\cdot$yr of atmospheric neutrino data at a DUNE-like detector, assuming IO. On the left is the sensitivity computed for fixed values of all other oscillation parameters for different values of the angular resolution below 1~GeV. 
On the right, we illustrate the effect of profiling over $\theta_{23}$ (green) and $\Delta m^2_{31}$ (gold), relative to the sensitivity obtained for fixed values of all oscillation parameters (blue curve), for a fixed value of the angular resolution below 1~GeV. }
\label{chi-eps_OMB_IO}
\end{figure}

\begin{figure}[htbp]
    \centering
    \includegraphics[width=0.49\linewidth]{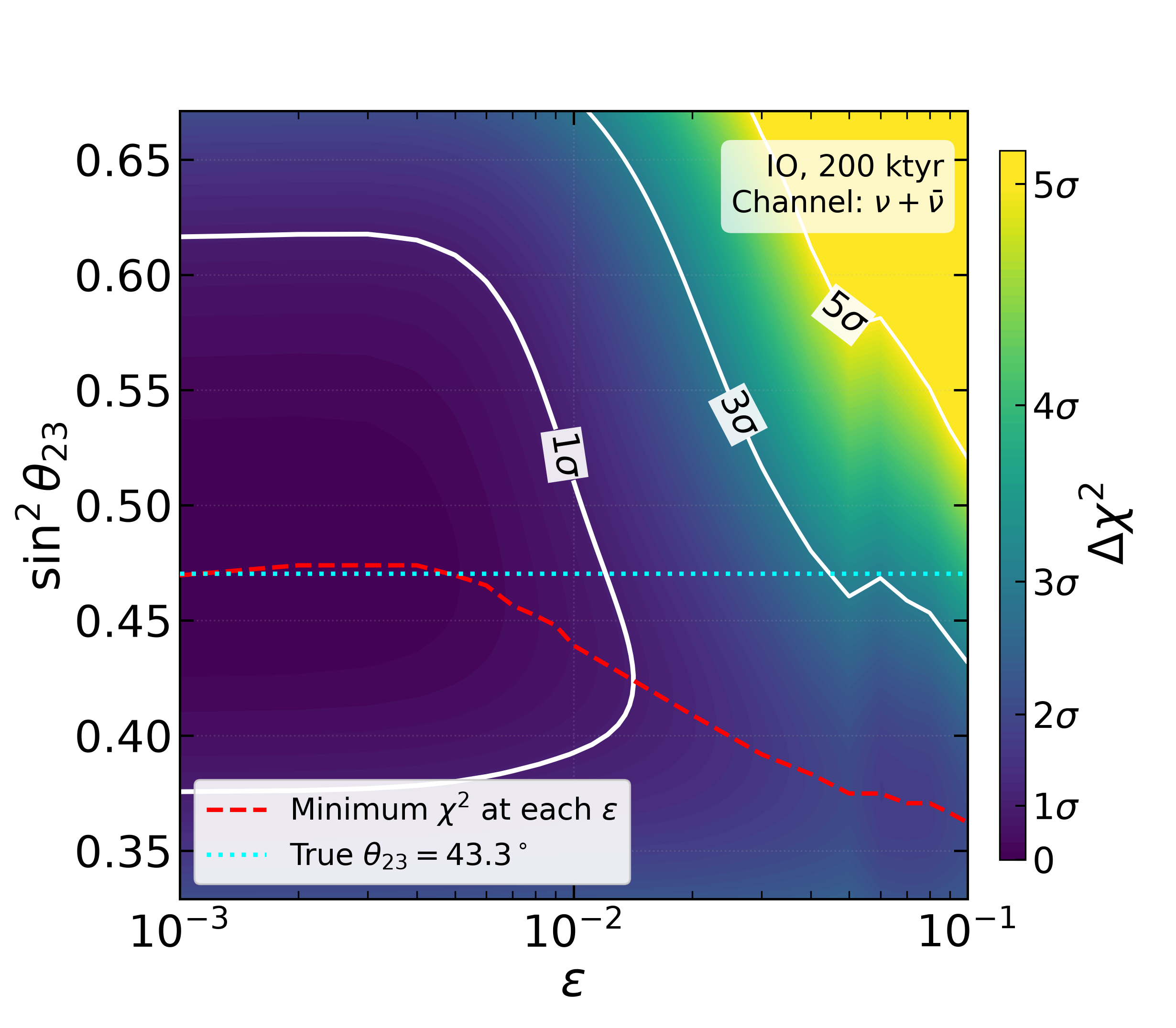}
    \includegraphics[width=0.49\linewidth]{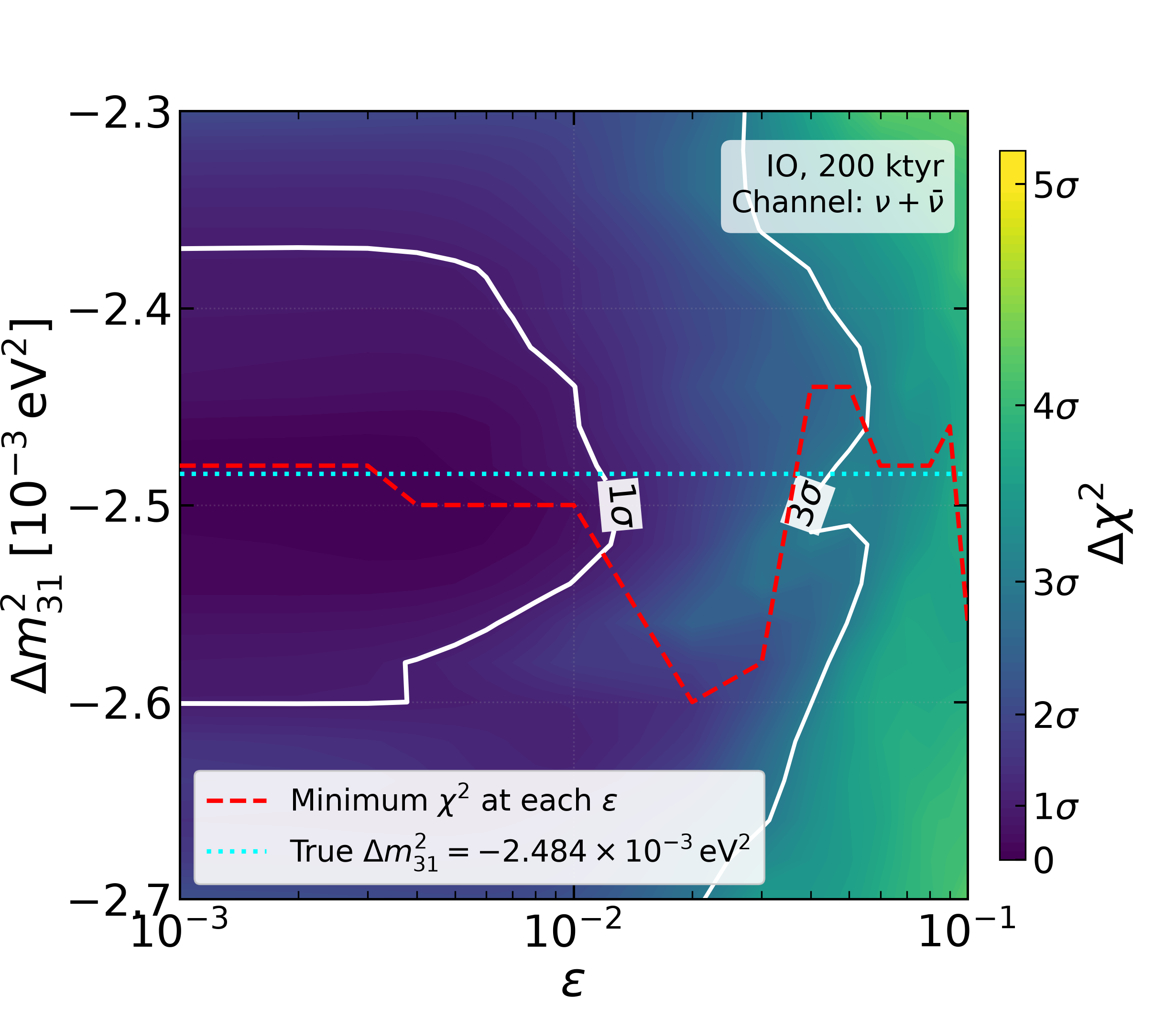}
    \caption{Sensitivity in the $\epsilon\times\sin^2\theta_{23}$ (left), and $\epsilon\times\Delta m^2_{31}$ plane (right) for IO. The red curve shows the position of the minimum value of $\chi^2$ as a function of $\epsilon$. The blue line indicates the true values on the relevant standard oscillation parameter.}
    \label{fig:chi-eps-sth23-dm231-io}
\end{figure}

\begin{figure}[htbp]
\centering
\includegraphics[width=0.45\columnwidth]{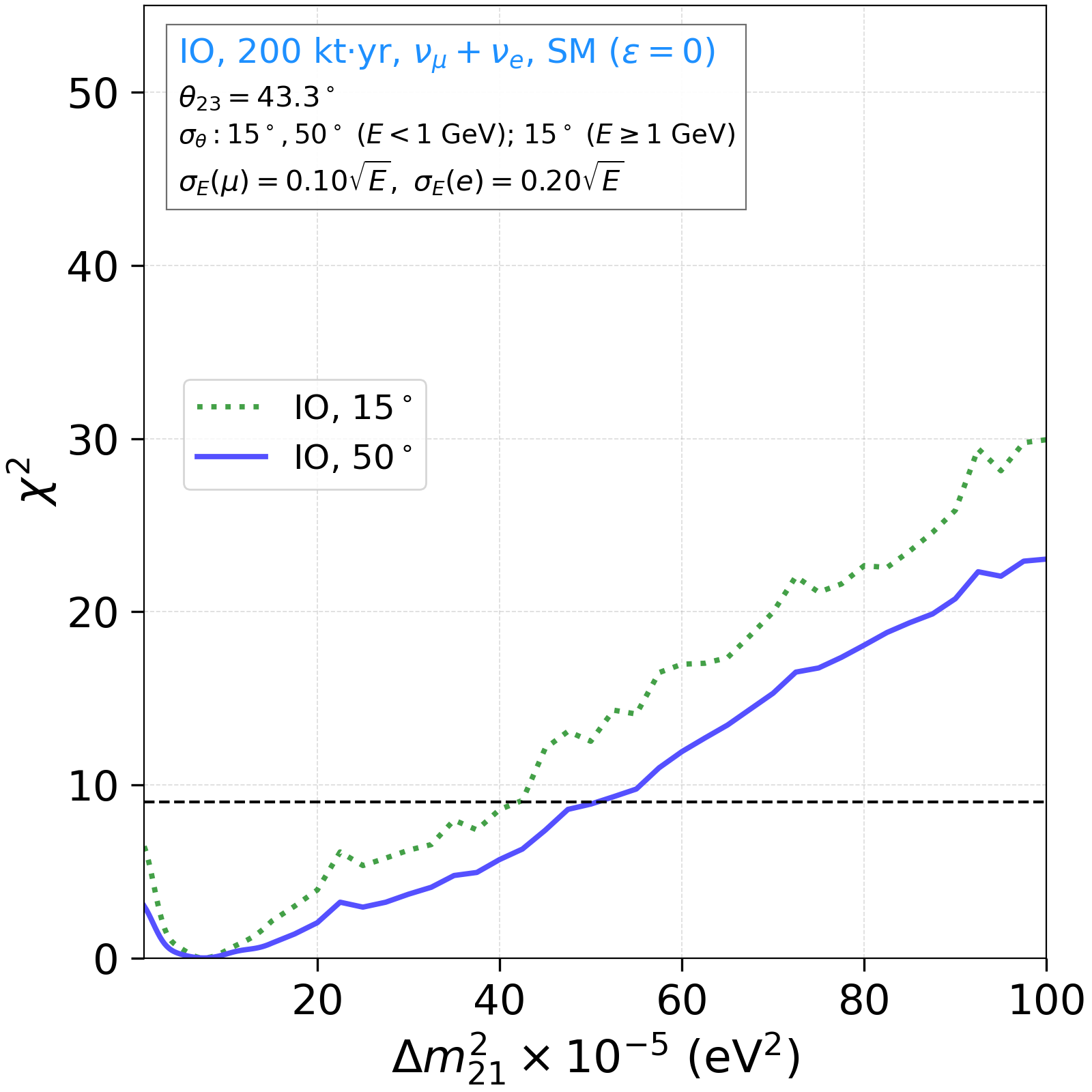}
\includegraphics[width=0.45\columnwidth]{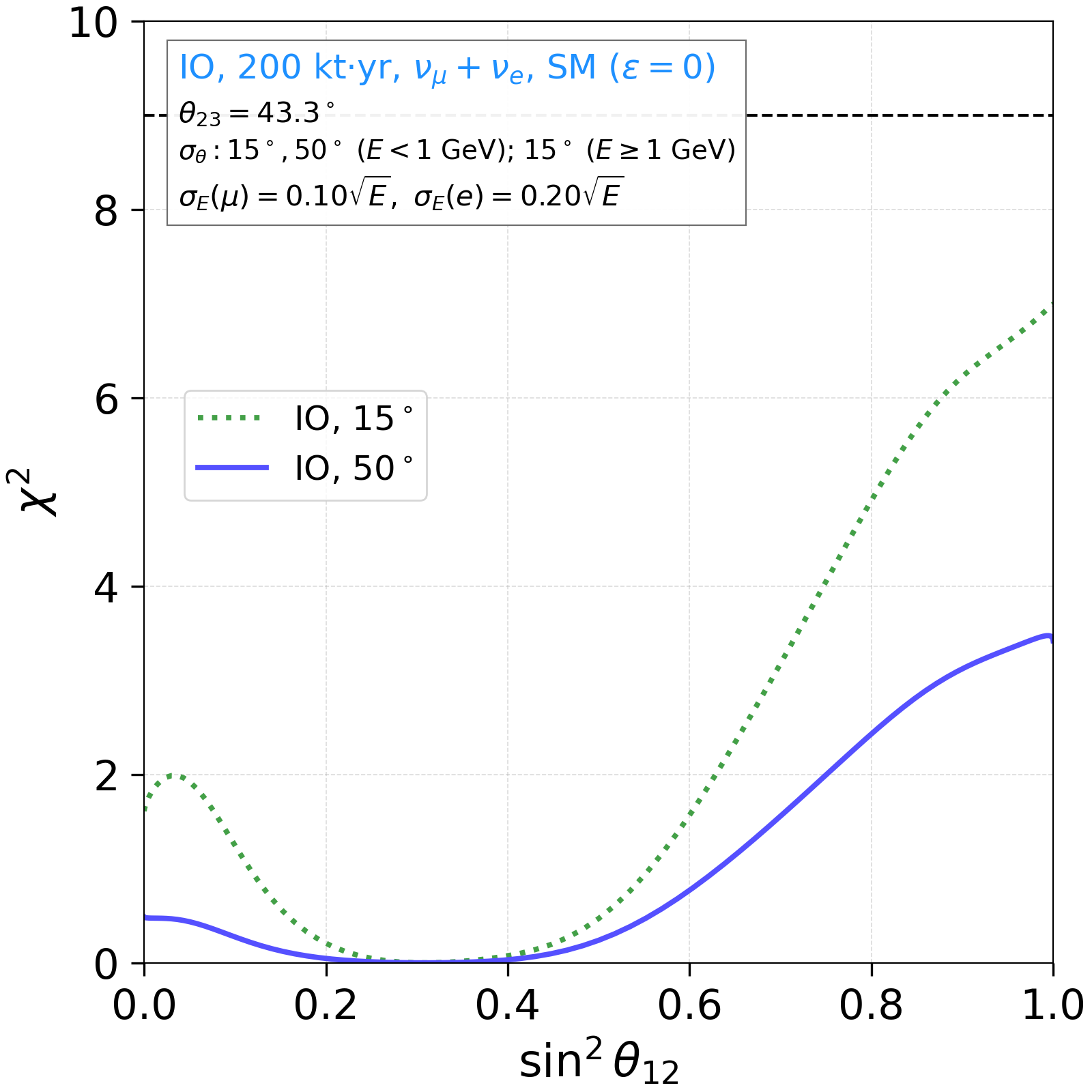}
\caption{Sensitivity to $\Delta m^2_{21}$ (left) and $\sin^2\theta_{12}$ (right) for standard oscillations,  assuming IO and 200 kt$\cdot$yr of data. The dotted and dashed correspond to different choices for the angular resolution when $E_\nu< 1$ GeV: $50^\circ$ (most conservative) and $15^\circ$ (most optimistic), respectively.}
\label{chi-eps-marg_OMB}
\end{figure}

\begin{figure}[H]
\centering

\includegraphics[width=0.48\columnwidth]{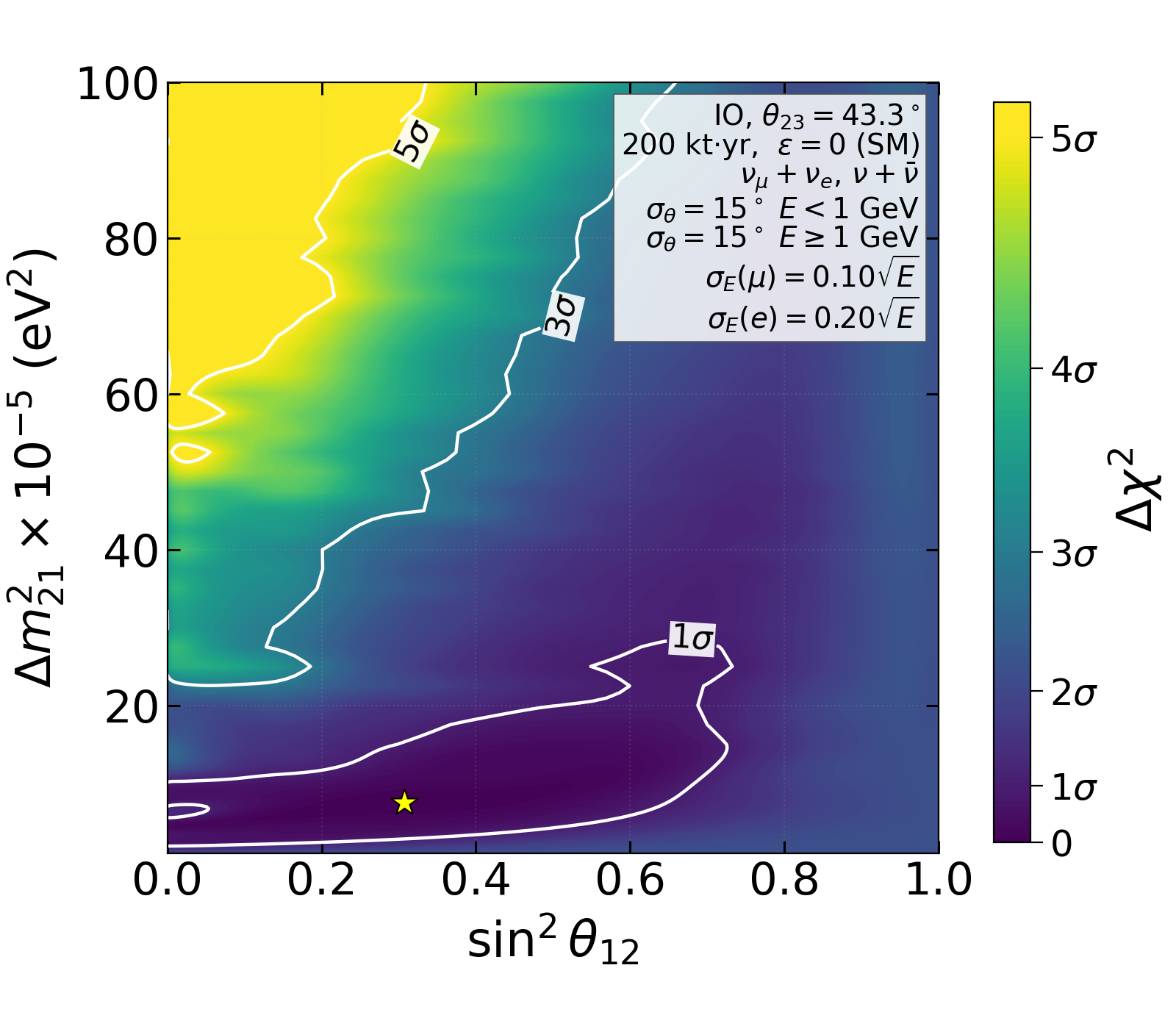}
\includegraphics[width=0.48\columnwidth]{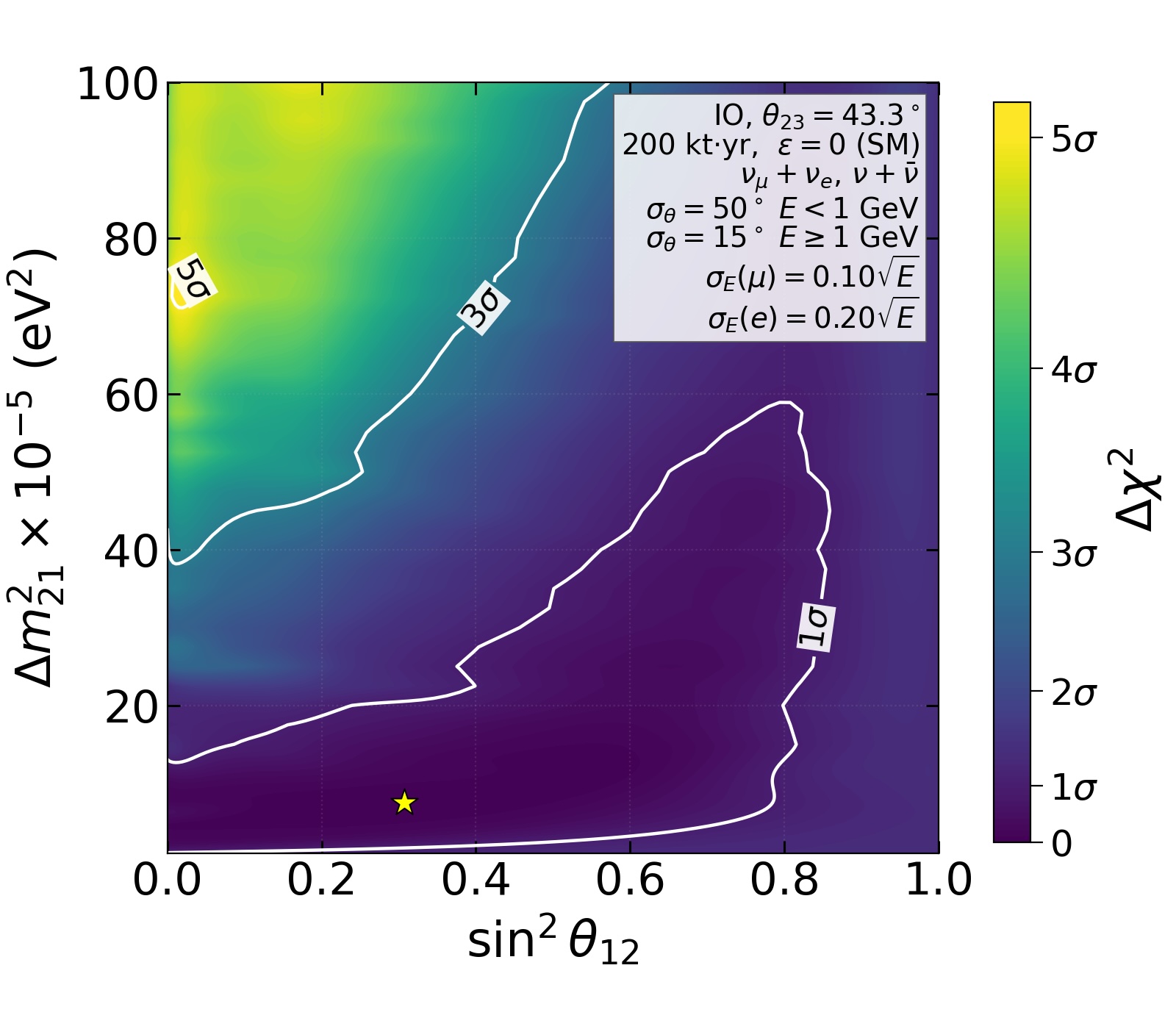}
\caption{Sensitivity to the solar oscillation parameters in the $\Delta m^2_{21}\times\sin^2\theta_{12}$-plane for standard oscillations assuming 200 kt$\cdot$yr of data. We assume IO. The left panel corresponds to an angular resolution for $E_\nu< 1$ GeV of $15^\circ$ (most optimistic), and the right panel corresponds to $50^\circ$ (most conservative).  The star in each panel indicates the best-fit value from NuFit \cite{Esteban_2024}.}
\label{dm21-th12-2d-sm_OMB_IO}
\end{figure}

\nocite{*}
\renewcommand{\bibsection}{\section*{References}}
\bibliography{apssamp_pseudo}

\end{document}